\documentclass[reprint,
superscriptaddress,
 amsmath,amssymb,
 aps,pra,
]{revtex4-2}

\usepackage{graphicx}
\usepackage{dcolumn}
\usepackage{bm}
\usepackage{esint}
\usepackage{xcolor}
\usepackage{anyfontsize}
\usepackage{lipsum}
\usepackage{svg}
\usepackage{url}
\usepackage{adjustbox}
\usepackage{booktabs}
\usepackage{physics}
\usepackage[bookmarksnumbered,pdfpagelabels=true,plainpages=false,colorlinks=true,linkcolor=blue,citecolor=blue,urlcolor=blue]{hyperref}
\usepackage{ulem}

\mathchardef\mhyphen="2D 

\begin{document}

\preprint{APS/123-QED}
\title{Toroidal Moments in Confined Nanomagnets and their Impact on Magnonics}

\author{F. Brevis}
\email{felipe.brevis@usm.cl}
\affiliation{Departamento de Física, Universidad Técnica Federico Santa María, Avenida España 1680, Valparaíso,
Chile}
\author{L. K\"orber}
\affiliation{Radboud University, Institute of Molecules and Materials, Heyendaalseweg 135, 6525 AJ Nijmegen, The Netherlands}
\author{B. Mimica-Figari}
\affiliation{Departamento de Física, Universidad Técnica Federico Santa María, Avenida España 1680, Valparaíso,
Chile}
\author{R. A. Gallardo}
\affiliation{Departamento de Física, Universidad Técnica Federico Santa María, Avenida España 1680, Valparaíso,
Chile}
\author{A. K\'akay}
\affiliation{Helmholtz-Zentrum Dresden-Rossendorf, Institute of Ion Beam Physics and Materials Research, Bautzner Landstr. 400, 01328 Dresden, Germany}
\author{P. Landeros}
\email{pedro.landeros@usm.cl}
\affiliation{Departamento de Física, Universidad Técnica Federico Santa María, Avenida España 1680, Valparaíso, Chile}

\date{\today}
             
\begin{abstract}
The nonreciprocity created by dipolar coupling, electric currents, and Dzyaloshinskii-Moriya interactions is discussed in cases where the magnon propagation direction has a component parallel to the toroidal moment. A criterion for calculating the toroidal moments is established, addressing the issue of correct origin selection by considering compensated and uncompensated magnetization distributions. This criterion is then applied to various nonreciprocal magnetic systems, with the calculations consistent with those reported in the literature and predicting the existence of nonreciprocity in a more general manner. These results broaden the physical significance of the toroidal moment and facilitate the identification and estimation of nonreciprocity in magnonic systems. 
This work also clarifies the interrelations between different definitions of the toroidal moment for confined structures, where a surface term arising from surface-bound currents connects these definitions without the need for time-averaging. Comparing these definitions of the toroidal moment applied to different magnetic textures demonstrates that they are always parallel but may differ in magnitude and sign. The discrepancy in the different definitions is deemed irrelevant since its direction, rather than its magnitude, primarily predicts the existence of magnon nonreciprocity. 
\end{abstract}

\maketitle

\section{\label{item1:intro}Introduction}
Nonreciprocity, the asymmetric propagation of waves or signals, is a fundamental concept across various physical domains, including photonics, phononics, plasmonics, electronics, spintronics, and magnonics  \cite{Camley87,Caloz18}. It is crucial for developing devices that control the flow of energy or information in a single direction, preventing unwanted reflections and improving system efficiency \cite{Kodera13,Peterson18}. For example, nonreciprocal optical isolators ensure that light travels in one direction while being blocked or altered in the opposite direction, which is vital for protecting optical systems \cite{Shi15}. In magnonics, nonreciprocity enables directional control of magnetic excitations known as spin waves (SWs), or magnons when described as quasiparticles, paving the way for quantum computing \cite{10.1063/5.0157520} and for technologies that enhance the performance and functionality of communication and data processing at the nanoscale \cite{An15,Chen22TopicalReview,Flebus24}.

Space-time symmetry breaking is an important condition for nonreciprocity. In magnetic materials, it can occur either from its interaction with quasiparticles, as photons interacting with magnetic crystals, or with magnons \cite{Cheong2018,Szaller13}. This condition can be indicated by the definition of a toroidal moment ($\boldsymbol{\tau}$) \cite{Gorbatsevich1994,Zimmermann14,Lehmann18}, allowing to estimate the presence of nonreciprocity based on the equilibrium magnetization ($\mathbf{M}$) and wave vector ($\mathbf{k}$).  
  It can be inferred from symmetry analysis that nonreciprocity is enabled if $\boldsymbol{\tau}\cdot \mathbf{k} \neq 0$, allowing for direction-dependent optical \cite{Szaller13} and magnetic \cite{Korber22} phenomena. Although this condition has already been reported, a deeper connection with nonreciprocity is still lacking. The presence of the toroidal moment $\boldsymbol{\tau}$ has been associated with the nonreciprocity of photons in a magnetic crystal  \cite{Foggetti2019,Mund2021}, and the nonreciprocity of spin waves in nanotubes with vortex state \cite{Korber22} and artificial chiral magnets with curling states \cite{Xu24}. Besides, such symmetry breakings are essential for the emergence of magnetoelectric effects, where the electric polarization ($\mathbf{P}$) couples with $\mathbf{M}$, as $\boldsymbol{\tau}$ is dual to the antisymmetric part of the magnetoelectric susceptibility tensor \cite{Spaldin08}. Thus, the presence of $\boldsymbol{\tau}$ is a key concept in condensed matter physics \cite{Marauska2012,Polcia2019,PourhosseiniAsl2020,Liang2021}.

\begin{figure}[!ht]
	\includegraphics[width=8.6cm]{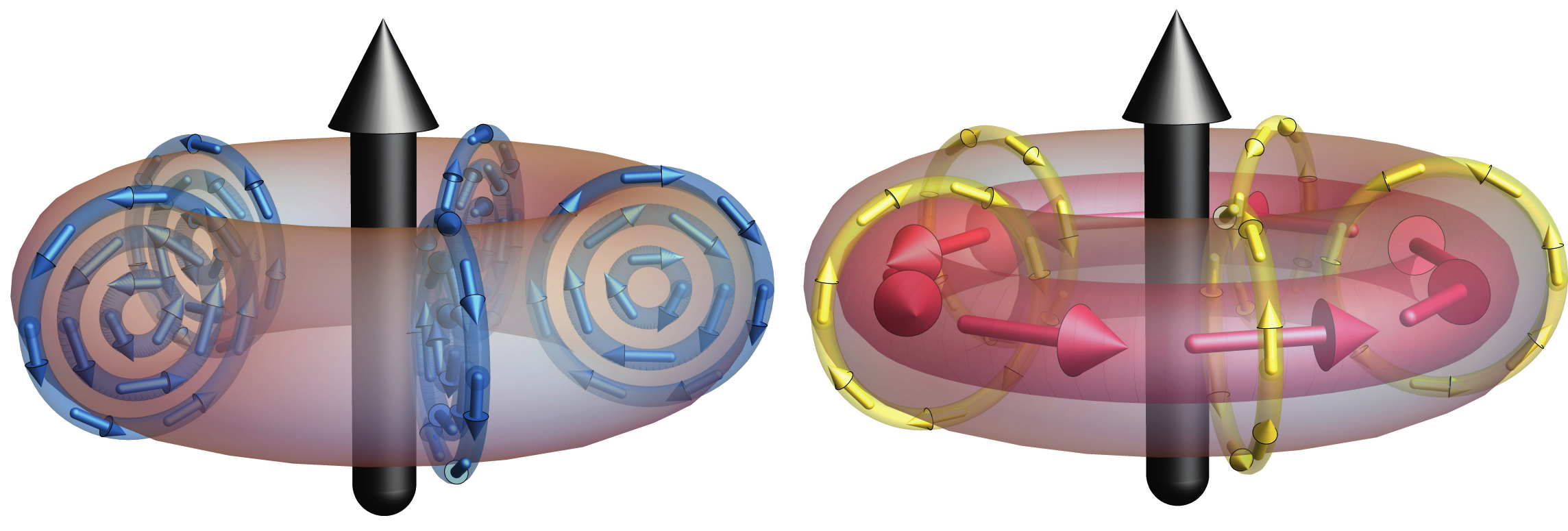}
	\caption{(Left panel) Toroidal moment (black central arrow) generated by any kind of circulating current distribution $\mathbf{J}$ (blue arrows). (Right panel) For bound currents only ($\mathbf{J}=\mathbf{J}_{\rm b}$) the toroidal moment has two contributions: a volumetric one, arising from the magnetization in the bulk (red arrows) and a surface contribution, arising from the surface bound currents (yellow arrows).}
	\label{fig:fig0}
\end{figure}

 Proposed from multipole expansions of the vector potential \cite{Dubovik90}, the polar toroidal moment arises as the lowest order term and can be understood considering an electric current moving along the meridians of a ferrite toroidal core, generating a magnetic field and a circular $\mathbf{M}$. As illustrated in Fig.~\ref{fig:fig0}, a toroidal moment perpendicular to the toroid plane is obtained \cite{Papasimakis16}. Microscopically, a toroidal moment $\boldsymbol{\tau}=(-g \mu_B/2)\sum_i \mathbf{r}_i \times \mathbf{S}_i$ can be viewed as a built-in effective vector potential $\mathbf{A}_\mathrm{eff}$ under the presence of spin–orbit interaction \cite{Tokura18}. In photonics, the effective spin-orbit coupling creates a toroidal moment along the electric field, allowing the observation of ferrotoroidic domains with second harmonic generation in LiCoPO$_4$ \cite{VanAken07,Zimmermann14}.
The same technique reveals the nonreciprocity of photons in a magnetoelectric antiferromagnet $\rm{CuB_{2}O_{4}}$ \cite{Mund2021}, which was first shown when the antiferromagnetic order vector is parallel to $\boldsymbol{\tau}$. It has also been demonstrated that the presence of $\boldsymbol{\tau}$, associated with the magnetoelectric effect, modifies the dispersion of electromagnetic waves in a medium, with a term proportional to $\boldsymbol{\tau} \cdot \mathbf{k}$ \cite{Kalish2007}, as well as the propagation of surface plasmons in nanowires \cite{Gusev2014}. In magnonics, a similar situation is described by Foggetti \textit{et al}. \cite{Foggetti2019}, studying magnons in hexagonal $\rm{LuFeO_3}$, showing that nonreciprocal spin-wave propagation is achieved when $\boldsymbol{\tau}\parallel\mathbf{k}$, both parallel to the crystal $c$-axis. From the symmetry of the lattice, the product $k_z ( \mathbf{r}_i \times \mathbf{S}_i)_z$ is invariant, and the magnon energy has a linear term proportional to $k_z$ (along the $c$ axis), giving rise to nonreciprocity \cite{Foggetti2019}. The essential point is that there is no symmetry operation connecting a reciprocal situation when $\boldsymbol{\tau}\parallel \mathbf{k}$  \cite{Cheong2018,Xu2024} because the symmetry operations will transform $\boldsymbol{\tau}$ and $\mathbf{k}$ in the same manner. 
Then, any frequency asymmetry in wave propagation arising from symmetry breaking can be anticipated if the condition $\boldsymbol{\tau} \cdot \mathbf{k} \neq 0$ is fulfilled \cite{Korber22}. This suggests that magnon nonreciprocity can be assessed by  $\boldsymbol{\tau}$, either if it arises from a free current , a curling magnetization \cite{Korber22,Xu24}, or from mirror symmetry broken by an interface (surface anisotropies, interlayer exchange coupling, or DMI) \cite{Matsumoto21}. Such phenomena are consequences of Neumann's principle, which states that any symmetry shown by the point group of the crystal is adopted by its physical properties \cite{Birss64,Landau,Szaller13}. 

In this paper, the implications of an emergent toroidal moment in the context of magnonics \cite{Barman21,Flebus24} are discussed and applied to a variety of systems, such as the current-induced SW Doppler shift via spin-transfer torque (STT) in films \cite{Lederer66,Fernandez04}, magnetic textures \cite{Ogawa2021,Seki2020,dosSantos20}, magnetization-graded films \cite{Gallardo19a}, bilayers and multilayers \cite{Gallardo19b,Sluka19,Albisetti19}, and films with interfacial and bulk Dzyaloshinskii-Moriya interaction (DMI) \cite{Cortes13,Gallardo19BCh,Kuepferling23,Matsumoto21}. First, the toroidal moment derived by Dubovik and Tugushev \cite{Dubovik90} is revisited using the concept of bound currents, showing that volume and surface toroidal contributions arise. 
The volume expression has been widely used to evaluate $\boldsymbol{\tau}$ \cite{Spaldin08,Talebi17}. Nonetheless, the surface term has not yet been reported and may be relevant for studying confined nanostructures. Finally, a theoretical approach, based on the Ederer and Spaldin model, is applied to predict symmetry breaking and nonreciprocity in magnonic systems. This model is used to avoid the origin dependence of $\boldsymbol{\tau}$ in microscopic systems \cite{Ederer07}.



\section{Toroidal Moment and Nonreciprocal Magnonics}

Several approaches for calculating $\boldsymbol{\tau}$ have been reported \cite{Spaldin08,Talebi17}. The first expression is \cite{Dubovik90}  
\begin{equation}
\boldsymbol{\tau} = \frac{1}{10}\int{dV\,\big[\mathbf{r}(\mathbf{r}\cdot\mathbf{J})-2r^2\mathbf{J} \big]},
\label{Eq1}
\end{equation}
where $\mathbf{J}$ corresponds to the current density, which may include free currents ($\mathbf{J}_{\rm{f}}$), bound currents ($\mathbf{J}_{\rm{b}}$), and polarization currents ($\partial \mathbf{P}/\partial t$) \cite{jackson1999classical,Griffiths_2017}. For magnetic matter, the bound current is $\mathbf{J}_{\rm{b}}=\nabla \times \mathbf{M}$. Then, in the absence of free and polarization currents, $\mathbf{J} = \mathbf{J}_{\rm{b}}$ and the magnetic contribution to the emergent toroidal moment is given by $\boldsymbol{\tau} = \frac{1}{10}\int{dV\,\big[\mathbf{r}(\mathbf{r}\cdot(\nabla \times \mathbf{M}))-2r^2\nabla \times \mathbf{M} \big]}$. By utilizing standard vector identities \cite{SuppMater}, $\boldsymbol{\tau}$ splits into volumetric and surface contributions $\boldsymbol{\tau} = \boldsymbol{\tau}^{\rm{v}} + \boldsymbol{\tau}^{\rm{s}}$, where
\begin{equation}\boldsymbol{\tau}^{\rm{v}}=\frac{1}{2}\int{dV\, (\mathbf{r}\times\mathbf{M})},
    \label{Eq2}
\end{equation}
which was later obtained by a multipole expansion of the vector potential \cite{Ederer07,Spaldin08}. Eq.~(\ref{Eq2}) has been widely used for calculating $ \boldsymbol{\tau}$ in magnetic structures \cite{Ding21,Zimmermann14,Lehmann18,Thle20,Cheong22}. Indeed, the microscopic version of the toroidal moment arises from Eq.~(\ref{Eq2}) \cite{Ederer07}. On the other hand,
\begin{equation}\boldsymbol{\tau}^{\rm{s}}=-\frac{1}{10}\oint_{\partial V}dS\big[\mathbf{r}\left(\mathbf{r}\cdot\mathbf{K}_{\rm{b}}\right)  -2r^2\mathbf{K}_{\rm{b}} \big] 
    \label{Eq3}
\end{equation}
is due to surface-bound current $\mathbf{K}_{\rm{b}}=\mathbf{M}\times \hat{n}$, with $ \hat{n}$ a unit vector normal to the local surface. Notice that $\boldsymbol{\tau}^{\rm{s}}$, which has not been reported before, is analogous to Eq.~\eqref{Eq1} but with a closed surface integral and opposite sign. Nonetheless, there is an alternative definition for the toroidal moment given by $ \boldsymbol{\tau}'=(1/6)\int{dV\, [\mathbf{r}\times(\mathbf{r}\times\mathbf{J})]}$ \cite{Spaldin08,Talebi17}, which gives $\boldsymbol{\tau}'=  (1/3)\boldsymbol{\tau}^{\rm{v}}-(5/3)\boldsymbol{\tau}^{\rm{s}}$, where $\boldsymbol{\tau}^{\rm{v}}$ and $\boldsymbol{\tau}^{\rm{s}}$ are the moments found before. The toroidal moment $\boldsymbol{\tau}$ and $\boldsymbol{\tau}'$ are equivalent when averaged over time if only free currents are considered \cite{Spaldin08}. Still, without this temporal average, there is a significative difference due to the surface term.

The toroidal moment can be used to predict the existence of dipolar symmetry breaking in the magnon spectrum \cite{Korber22}. However, one issue that is not considered in the relation of the toroidal moment with spin dynamics is the proper choice of origin when calculating the volume toroidal moment. One can easily check, e.g., with Eq.~\eqref{Eq2}, that shifting the coordinate origin by a vector $\mathbf{R}$ changes the toroidal moment $\boldsymbol{\tau}^{\rm v}\mapsto \boldsymbol{\tau}^{\rm v} + (1/2)\,V\, \mathbf{R}\times \langle \mathbf{M} \rangle$, where $V$ is the sample volume. Nonetheless, the choice of origin is not an issue if the sample has a zero magnetic moment, $\langle \mathbf{M}\rangle=0$, such as the vortex texture \cite{Korber22}.
One route to circumvent this origin dependence is found in Ref. \citenum{Ederer07} for localized spins, where $\mathbf{M}$ is divided into a compensated part with zero magnetic moment $\mathbf{M}^{(0)} = \mathbf{M} - \langle \mathbf{M} \rangle$ and an uncompensated or fully polarized background $\mathbf{M}^{(1)} = \langle \mathbf{M} \rangle$. Following Ref.~\citenum{Ederer07}, the part of the toroidal moment due to $\langle \mathbf{M} \rangle$ cannot induce any frequency asymmetry for a homogeneous and centrosymmetric lattice (in the absence of intrinsic DMI). Therefore, the only part of the toroidal moment that is relevant for dipolar symmetry-breaking is \footnote{From this point forward, $\boldsymbol{\tau}^{\rm v}$ will refer to Eq. (\ref{Eq4}), a more general formulation compared to Eq. (\ref{Eq2}).}
\begin{align}
    \boldsymbol{\tau}^{\rm v}  = \frac{1}{2}\int{dV\, \mathbf{r} \times (\mathbf{M} - \langle \mathbf{M} \rangle), }
    \label{Eq4}
\end{align}
which is origin-independent, as it involves only the compensated magnetization. 
This expression proves especially useful, as will be demonstrated in the following sections, for nonsymmetric structures such as partially closed tubes (Subsection~\ref{subsec:texture_curved_shells}), as well as for systems with spatial variation in $M_{\rm s}$, including graded films and multilayers with differing thicknesses and saturation magnetizations (Subsection~\ref{subsec:graded_multilayers}).
While $\boldsymbol{\tau}$ can be estimated using Eqs.~(\ref{Eq1}-\ref{Eq4}), in general, their different definitions are parallel to each other \cite{SuppMater}. Then, $\boldsymbol{\tau} \cdot \mathbf{k}$ can be calculated using any of these definitions.
In the following examples, the different definitions of the toroidal moment are used to analyze spin-wave nonreciprocity in different magnetic architectures. 
Here, it is emphasized that Eq.~(\ref{Eq1}) is suitable to connect the toroidal moment with nonreciprocal wave phenomena when only free currents are present ($\mathbf{J} = \mathbf{J}_{\rm{f}}$). 
In the absence of free currents, and for nonuniform magnetization textures, Eqs.~(\ref{Eq3}-\ref{Eq4}) are used to capture the proper relation between the nonreciprocal wave propagation and the toroidal moment \footnote{In the following, the symbol $\boldsymbol{\tau}$ will be used to denote any kind of toroidal moment unless it is necessary to specify a volume or a surface toroidal moment.}.

Nonreciprocal propagation is usually described through the frequency shift among counterpropagating spin waves $\Delta f = f(\mathbf{k})-f(-\mathbf{k})$. Analytical formulae for $\Delta f$ have been obtained only for some magnetic systems \cite{Cortes13,Gallardo19b,Fernandez04,Otalora16,Gallardo22}.
A general expression can be derived from Ref.~\onlinecite{Cortes13}, which reads 
\begin{equation}
    \Delta f=\frac{\gamma \mu_0M_{\rm s}}{\pi}{\rm Im}[\mathcal{N}^{(21)}(\mathbf{k})],
\label{Eq-N21}
\end{equation}
where $\gamma$ is the absolute value of the gyromagnetic ratio, $\mu_0$ is the vacuum permeability, $M_{\rm{s}}$ is the saturation magnetization, and $\mathcal{N}^{(21)}(\mathbf{k})$ is the off-diagonal element of the spin-wave tensor in a local basis, which has the property $\mathcal{N}^{(12)}=[\mathcal{N}^{(21)}]^*$ \cite{Korber21b}. It is worth emphasizing that such a formula is not restricted to small $k$, and may lead to nonlinear wave-vector dependence due to the dipolar coupling \cite{Korber21b,Gallardo22,KorberT23,Mimica25SW}. 
If $\Delta f$ is expanded up to first order in the wave vector, then it can be shown that ${\rm Im}[\mathcal{N}^{(21)}]\propto (\boldsymbol{\tau} \cdot  \mathbf{k})$. Namely, the toroidal moment allows for estimating nonreciprocity in the small $k$-limit.

\subsection{Current-induced nonreciprocity}
By considering a constant electric current density $\mathbf{J}_{\rm{f}}$ flowing through a film with volume $V$, one finds, from Eq.~\eqref{Eq1}, $\boldsymbol{\tau} \propto -V \mathbf{J}_{\rm{f}}$. Consequently, the nonreciprocity condition gives $\boldsymbol{\tau} \cdot  \mathbf{k} \propto -\mathbf{J}_{\rm{f}} \cdot \mathbf{k}$, which is nonzero if the wave vector includes a component along the current. This statement aligns perfectly with earlier works. Indeed, an electric current, via spin-transfer torque, produces a linear term in the magnonic dispersion associated with a Doppler shift $\Delta f \propto \mathbf{J}_{\rm{f}} \cdot \mathbf{k}$, which has been predicted by theory \cite{Lederer66,Fernandez04} and measured experimentally \cite{Vlaminck08,Zhu10} for a saturated magnetic film. Therefore, a free electric current creates a toroidal moment and a nonreciprocal wave behavior, even in the case of uniform magnetization and regardless of its orientation. In the case of nonuniform current densities (not shown), it can be demonstrated that the toroidal moment is still parallel to the current, which is also the case for surface current densities. 
On the other hand, it can be shown that the imaginary part of the spin-transfer torque magnetic tensor (due to an electric current $\mathbf{J}_{\rm{f}}$) is given by \cite{GallardoUnp} 
\begin{equation}
{\rm{Im}}[\mathcal{N}_{\rm stt}^{(21)}(\mathbf{k})]=\frac{\mu_{\rm B} \mathcal{P}}{\gamma\mu_0 M_{\rm s}^2 e}  \left(\frac{1+\alpha\beta}{1+\beta^2}\right) (\mathbf{J}_{\rm{f}} \cdot  \mathbf{k}),
\end{equation}
which shows that the STT frequency nonreciprocity is linear in the wave vector \cite{Lederer66,Fernandez04,Vlaminck08}. Here, $\alpha$ is the Gilbert damping, $\beta$ is the nonadiabatic STT term, $\mu_{\rm B}$ the Bohr magneton, $e$ the absolute value of the electron charge, and $\mathcal{P}$ the spin polarization. The adiabatic and nonadiabatic contributions to the effective field, for low currents, are proportional to $\mathbf{M}\times(\mathbf{J}_{\rm f}\cdot\nabla)\mathbf{M}$ and $(\mathbf{J}_{\rm f}\cdot\nabla)\mathbf{M}$, respectively. These STT effective fields do not accurately capture the dynamics for higher currents.  As $\boldsymbol{\tau}\propto \mathbf{J}_{\rm f}$, it can be concluded that $\Delta f_{\rm stt}\propto (\boldsymbol{\tau}\cdot \mathbf{k})$, which is always linear in $k$ for low current densities.

\begin{figure}[b]
	\includegraphics[width=8.6cm]{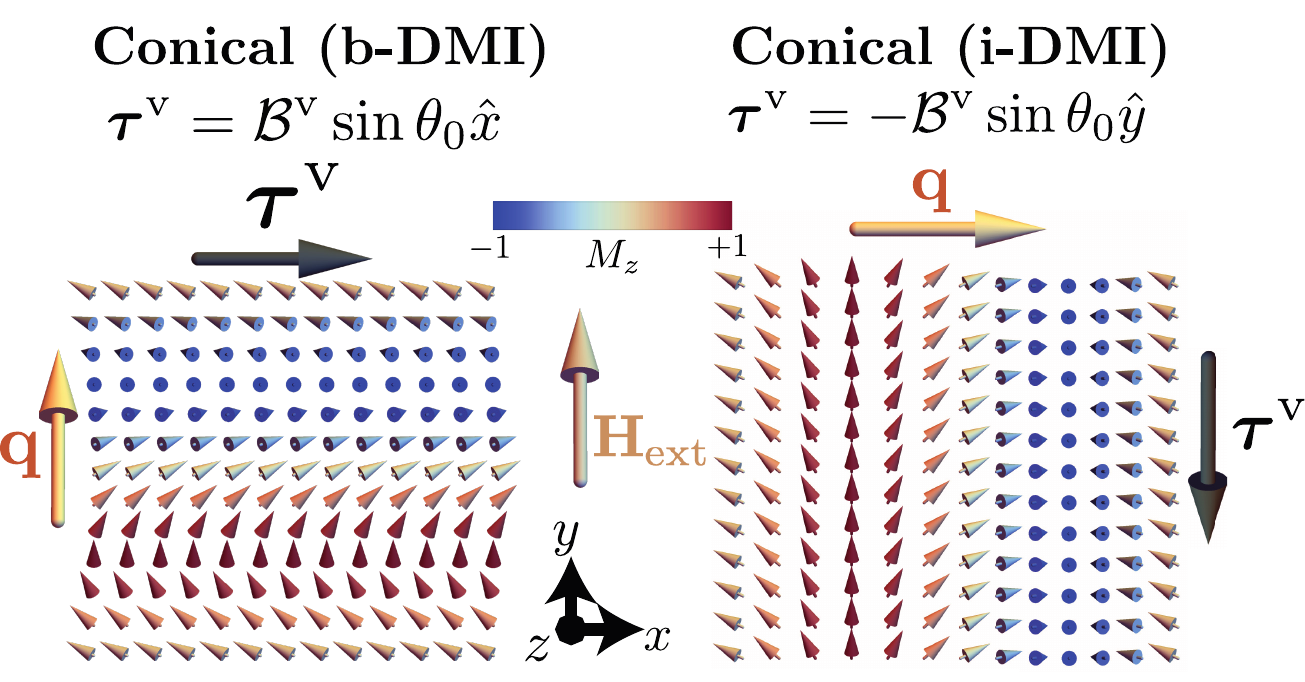}
	\caption{Conical-helix textures in a thin film with bulk (left) and interfacial (right) DM interaction, where the applied field $\mathbf{H}_{\rm ext}$ is along $y$ such that the cone-angle is $\theta_0 = \pi/3$ and the phase is $\psi=\pi/2$. The in-plane helical pitch vector $\mathbf{q}$ is parallel (perpendicular) to the field for bulk (interfacial) DMI. The resulting volume toroidal moment (black arrow) is illustrated in both cases and is always in the plane and perpendicular to the pitch vector. }
	\label{fig:ch_textures}
\end{figure}

\subsection{Texture-based nonreciprocity in flat films}
In the absence of free currents but in the presence of noncollinear forms of the magnetization that may exhibit helicity or chirality \cite{Yu2012,Hyu21,Yu23}, a toroidal moment should emerge from a bound current. Examples are conical-helix (Bloch or Néel type) \cite{Kugler15,Schwarze15,Weiler17,Garst17,Rios22}, vortex-like \cite{Korber22}, skyrmion-like \cite{Rowland16,Gbel2019,Bhowal2022}, and hopfion-like magnetic textures \cite{Rybakov22,Saji23}. 
A conical-helix (CH) texture (see Fig.~\ref{fig:ch_textures}), stabilized by a DMI, can be modeled as $\mathbf{M}(\mathbf{r})= M_{\rm{s}}\big( \sin[\mathbf{q}\cdot \mathbf{r}+ \psi] \sin \theta, \cos \theta,  \cos[\mathbf{q}\cdot \mathbf{r}+ \psi] \sin \theta  \big)$, where $\mathbf{q} =q(\sin\varphi_{\mathbf{q}},\cos\varphi_{\mathbf{q}},0)$ is the in-plane pitch vector that establishes the direction of the texture variation according to $\varphi_{\mathbf{q}}$ \cite{Rios22}. 
For bulk DMI (b-DMI), the angle $\varphi_{\mathbf{q}}=0$ produces a Bloch-like texture. Instead, for interfacial DMI (i-DMI), $\varphi_{\mathbf{q}}=\pi/2$ induces a Néel-like texture. The phase angle $\psi$ must be accounted for finite systems and is related to the tendency of $\mathbf{M}$ to orient in-plane or out-of-plane due to the competition between the dipolar field and perpendicular anisotropy \cite{Rios22}. Also, $\theta$ is the cone angle related to the tilting around the external field ($\mathbf{H}=H\hat{y}$), so that $\theta=\pi/2$ describes a perfect helix, while $\theta=0$ defines the polarized state along the field. The toroidal moment for the conical-helix, either with bulk or interfacial DMI, gives $\boldsymbol{\tau}^{\rm{v}}= \mathcal{B}^{\rm{v}}\sin\theta(\sin\psi\cos\varphi_{\mathbf{q}},-\sin\psi \sin\varphi_{\mathbf{q}},\cos\psi \cos\varphi_{\mathbf{q}})$,
where $\mathcal{B}^{\rm{v}} = \frac{M_{\rm{s}} Ld}{2q^2}  \left[L q \cos \left(Lq/2\right)-2 \sin \left(Lq/2\right)\right]$, considering a square film with volume $V = L^2 d$ and thickness $d$.
For this texture, $\boldsymbol{\tau}^{\rm{v}}$  is always parallel or antiparallel to $\boldsymbol{\tau}^{\rm{s}}$ \cite{SuppMater}. Fig.~\ref{fig:ch_textures} illustrates the volume toroidal moment directions for conical-helix magnetic textures. These results and the nonreciprocity condition are consistent with Ref.~\citenum{Ogawa2021}, where asymmetric dispersions were reported for SWs excited over CH states in a Cu$_2$OSeO$_3$ single crystal with bulk DMI. Here, a frequency shift was observed for spin waves with a wavevector parallel to  $\boldsymbol{\tau}^{\rm v}=\tau^{\rm v}_{x} \hat{x}$ (and perpendicular to $\mathbf{H}$) \cite{Ogawa2021}, which is not the case for uniformly magnetized films with bulk DMI, where nonreciprocity occurs for $\mathbf{k}\parallel\mathbf{H}$  \cite{Cortes13}. The difference arises solely from the texture. 
It is worth mentioning that no analytical expressions currently exist for the frequency nonreciprocity or the effective magnetic tensor associated with the conical-helical texture, making the toroidal moment estimation very relevant for determining the presence or absence of nonreciprocity arising from spin-helical magnetic textures.

The previous discussion focuses on a specific periodic magnetic texture. Nevertheless, the analysis of the toroidal moment and its connection with the nonreciprocity also applies to an arbitrary periodic texture.
In this case, the magnetization can be expanded using a Fourier series in terms of reciprocal vectors $\mathbf{G}$, where it is found that the toroidal moment is always perpendicular to $\mathbf{G}$. 
This can have important applications in the study of multi-$\mathbf{G}$ spin textures, such as skyrmion lattices \cite{Muhlbauer09}, multiple-$q$ spin textures \cite{Takagi18, Fujishiro19,Hayami20c}, tetrahedral-$q$ hedgehog and cubic-3-$q$ hedgehog lattices.
In those cases, analyzing the toroidal moment associated with each periodicity vector $\mathbf{G}$ offers a straightforward way to identify nonreciprocal wave directions as well as the possibility of studying ferrotoroidal orders in molecular systems \cite{Zimmermann14,Vignesh17} and artificial spin ices \cite{Lehmann18,Yue24,Rodriguez23}.

\begin{figure}[!ht]
	\includegraphics[width=8.6cm]{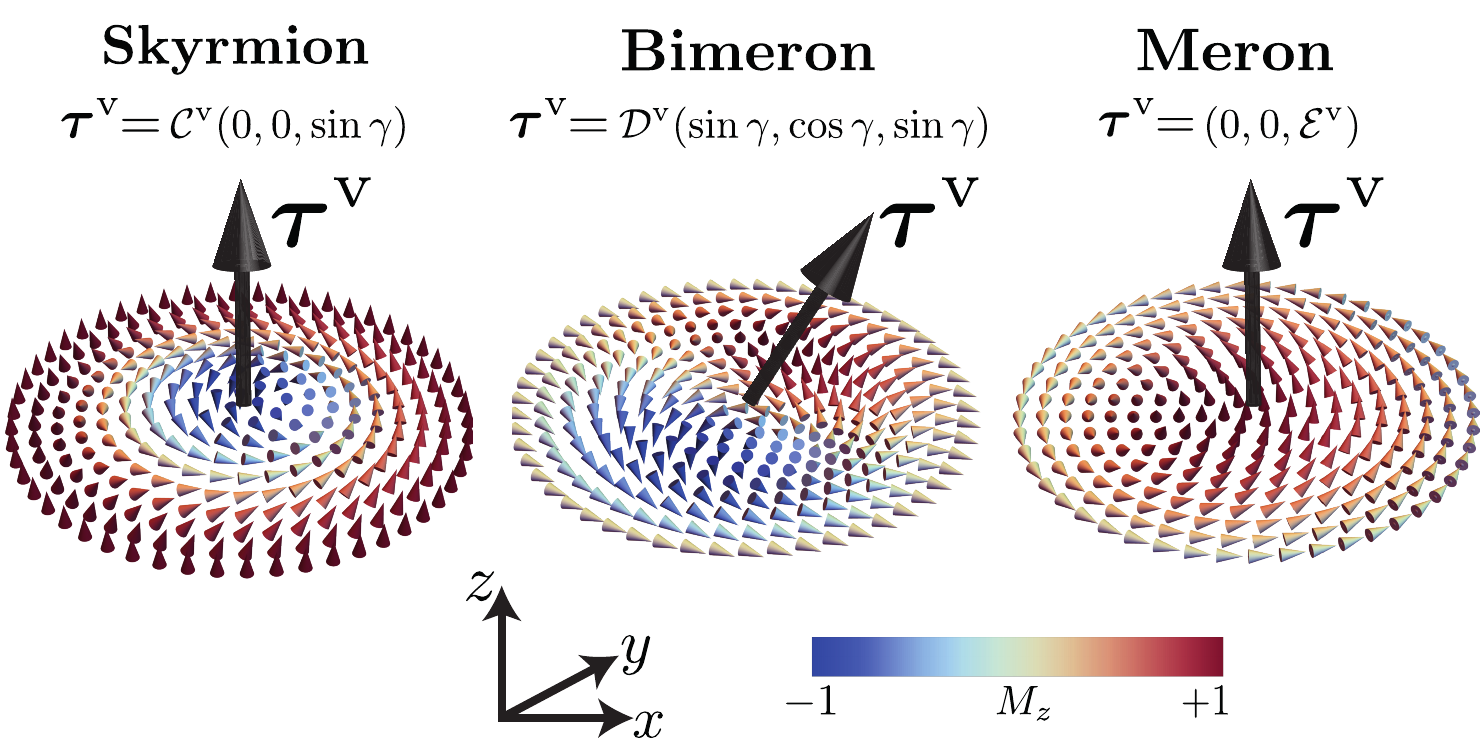}
	\caption{Skyrmion, bimeron, and meron magnetic textures. Coloring according to the $z$-component and the helicity is $\gamma=\pi/2$ for skyrmion and bimeron. The functions $\mathcal{C^{\rm{v}}}$, $\mathcal{D^{\rm{v}}}$, and $\mathcal{E^{\rm{v}}}$ can be found in the supplementary material \cite{SuppMater}. For each texture, the black arrow illustrates the calculated volume toroidal moment. }
	\label{fig:sk_textures}
\end{figure}

Skyrmionic textures in confined nanostructures also produce a $\boldsymbol{\tau}^{\rm v}$ pointing out of the plane for the vortex, meron, skyrmion, and bimeron textures, with the latter also exhibiting an in-plane component. Fig.~\ref{fig:sk_textures} illustrates the volume toroidal moments based on analytical models for a disk with radius $R$ and thickness $d$ (see supplementary material \cite{SuppMater} for more details), where the helicity $\gamma = 0\,(\pi/2)$ corresponds to a Néel (Bloch) skyrmion. As an extension, it is also possible to model skyrmioniums/bimeroniums or $\ell\pi$-skyrmions/bimerons, where $\ell=1,2,3...$ and $R = \ell\lambda$, with $\lambda$ being the texture period along the radial direction. These magnetic textures are composed of a combination of skyrmions or multiple full spin rotations along the diameter with a different topological charge \cite{Pepper2018,Cortes2019,Mehmood2021,Ponsudana2021}. These toroidal moment calculations are consistent with those observed for skyrmion strings \cite{Seki2020,Xing20}, where the magnetization induces a toroidal moment parallel to their axis, and the nonreciprocal wave effect is produced along such an axis \cite{Seki2020}. Importantly, the toroidal moments in antiskyrmions are always zero. From these results, one can expect similar behavior for SWs propagating along bimeron-like structures such as strings or lattices since their toroidal moment, depending on $\gamma$, can point in any of the three main directions. Similar to the conical-helix texture, no analytical expressions are available for $\Delta f$, which makes the toroidal moment calculation very relevant for addressing the presence of nonreciprocity in skyrmionic textures.

\subsection{Texture-based nonreciprocity in curved shells}
\label{subsec:texture_curved_shells}

 Another exciting example is magnon propagation along the axis of nanotubes with magnetization in vortex or curling states, where asymmetric dispersions originating from the magnetostatic interaction have been predicted \cite{Otalora16,Sheka20} and measured \cite{Korber21}. This is evident when analyzing the magnetization texture of the curling state, which in cylindrical basis $(\hat{\rho},\hat{\phi},\hat{z})$, is given by $\mathbf{M} =  M_{\rm s} \big(0, \chi\sin\Theta,\cos\Theta \big)$, where $\chi$ is the helicity ($\pm 1$) around the $z$-axis, and $\Theta$ the constant angle between the equilibrium magnetization and the $z$-axis. 
 Such state produces  $\boldsymbol{\tau}^{\rm v} = \frac{\pi}{3}(R^3-R_i^3)L\chi M_{\rm s} \sin\Theta\, \hat{z}\propto M_{\phi}\hat{z}$, where $R_i(R)$ is the internal(external) radius and $d=R-R_i$ is the shell thickness. For a vortex configuration,  $\Theta=\pi/2$, and 
 $\boldsymbol{\tau}^{\rm v}$ is parallel to the tube axis \cite{Korber22}, explaining the emergence of dipolar nonreciprocity along the axis. In contrast, for axial magnetization, $\Theta=0$ and $\boldsymbol{\tau}^{\rm v}=\boldsymbol{\tau}^{\rm s}=0$, in agreement with previous works \cite{Gallardo22,Korber22}. The curling magnetization in a closed nanotube is a very special case since there are analytical solutions available for the frequency nonreciprocity which incorporates dipolar \cite{Otalora16,Sheka20,Gallardo22,Korber22}, dipolar and exchange \cite{SalazarCardona2021}, and dipolar, exchange and DMI \cite{Mimica25SW,KorberT23}.
In the latter case, due to curvature, the spin waves are characterized by an axial wave vector $k_z$ and a quantized azimuthal wave vector proportional to the mode-index $l=0,\pm1,...$, that is $\mathbf{k} = k_z \hat{z} + (l/R)\hat{\phi}$. 
Then, the relevant magnetic tensors [see Eq.~\eqref{Eq-N21}], associated with each energy term, can be obtained, which leads to three contributions to the frequency asymmetry
\begin{equation}
    \Delta f^{\rm dip} = -\frac{\mu_0 \gamma}{\pi}\left(R_{\rm m} k_z M_\phi - R k_\phi M_z \right)\,\mathcal{I}_{1}^{k_z,l}, 
\end{equation}
\begin{equation}
    \Delta f^{\rm{ex}} = -\frac{\mu_0 \gamma}{\pi} k_\phi M_z \left(\frac{2 \ell_{\rm{ex}}^2}{R_i} \right),
\end{equation}
and 
\begin{equation}
   \Delta f ^{\rm dm}= \frac{\mu_0 \gamma }{\pi} \left(\frac{2D}{\mu_0 M_{\rm s}^2}\right) \left(k_z  M_\phi-k_\phi M_z\right)
\end{equation}
respectively associated with the dipolar, exchange, and interfacial Dzyaloshinskii-Moriya interactions. Here, $R_{\rm m}=(R+R_i)/2$ is the average radius, $D$ is the DM strength, $\ell_{\rm{ex}}$ is the exchange length and $\mathcal{I}_{1}^{k_z,l}$ is an azimuthal integral arising from dipolar coupling that is even in $k_z$ and also depends on the mode-index $l$ \cite{Mimica25SW,KorberT23}.
For the purposes of the present article, only dipolar and DM interactions are relevant due to the fact that the linear toroidal moment is a global property and cannot account for the exchange-induced nonreciprocity from curvature, nor a circulating toroidal moment. 
In the small $k_z$ limit, the leading term in $\mathcal{I}_{1}^{k_z,l}$ is a constant, explaining the linear $\Delta f$. 
Interestingly, in the thin-shell limit,
\begin{align}
    \Delta f^{\rm dm+dip}   
     = - \frac{\mu_0 \gamma} {\pi}\qty(\frac{2D}{\mu_0 M^2_{\rm s}}+\frac{d}{2}\delta_{l,0} ) (\mathbf{M}\times \hat{n})\cdot \mathbf{k},
\end{align}
where $ \hat{n}=\hat{\rho}$ is the normal to the cylindrical shell. 
It is important to mention that higher-order contributions [$\mathcal{O}({k_z^2})$] in $\mathcal{I}_{1}^{k_z,l}$ arise for larger $k_z$ \cite{Mimica25SW,KorberT23}, which leads to a nonmonotonic $\Delta f(k)$ typical of the nonlocal dipolar coupling, which is the only energy term that produces a nonlinear term in $\Delta f(k_z)$. 
In the small $k_z$ limit, and for ultrathin shells, the dipolar and DMI contributions are proportional to $\mathbf{K}_b\cdot\mathbf{k}$, where $\mathbf{K}_b=\mathbf{M}\times\hat{\rho}$ is the bound surface current in the outer cylindrical shell, leading to a surface toroidal moment proportional to $\mathbf{M}\times \hat{n}$ \cite{Kuznetsov2022}.


\begin{figure}[!ht]
    \includegraphics[width=8.8cm]{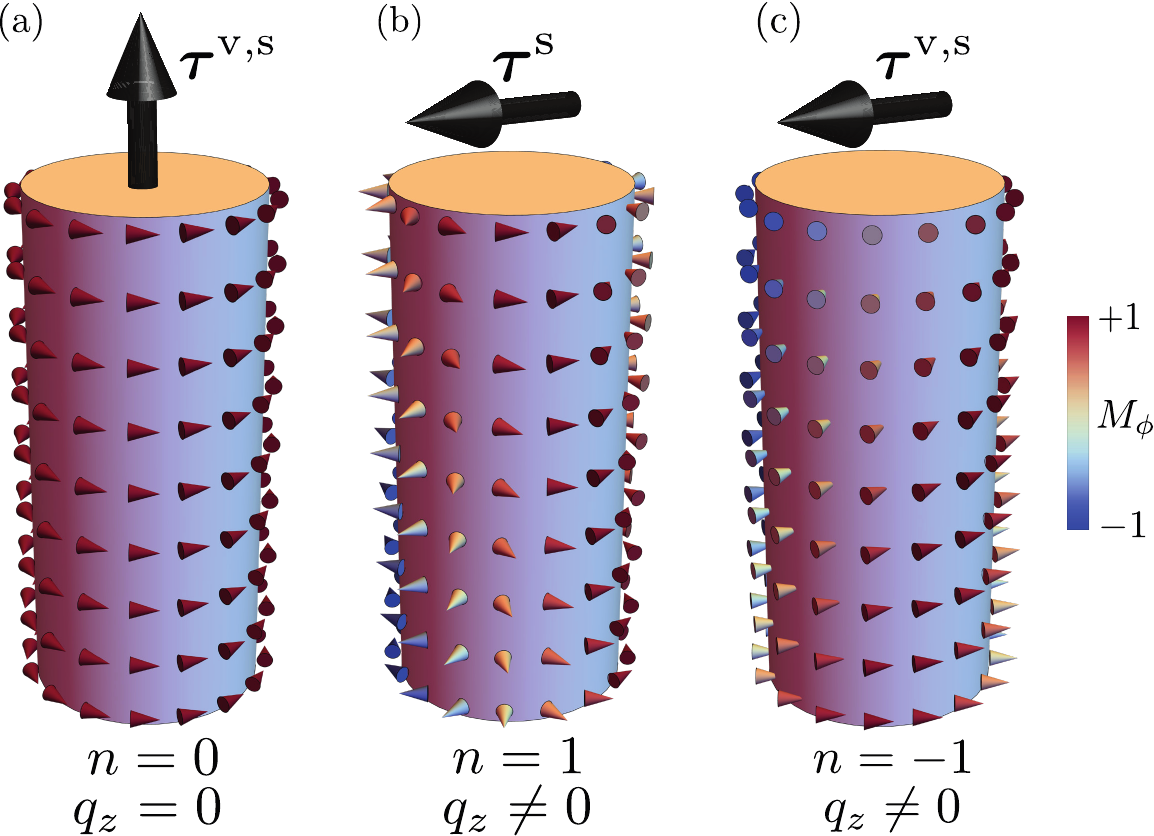}
	\caption{Illustration of three possible conical-helix magnetic textures on a nanotube described by Eq.~\eqref{eq:conical-helix-tube}. (a) Vortex state with $n=0$, $q_z=0$, and $\psi=\pi/2$, $\boldsymbol{\tau}$ points along the axis. Conical-helix state with (b) $n=1$ and (c) $n=-1$, both with finite $q_z$, arbitrary $\psi$, resulting in toroidal moments transversal to the axis. Coloring according to the azimuthal direction of magnetization.
}
	\label{fig:tubes_dmi_textures}
\end{figure}

Beyond the previously discussed curling state, conical-helix textures are also expected to emerge in ultrathin nanotubes with interfacial DMI similar to what has been observed in planar films \cite{Yershov20,Mimica25a}.
However, in contrast to flat films, closed tubular systems exhibit quantized magnetic textures along the azimuthal direction, characterized by an azimuthal component of the pitch vector $q_{\phi}\propto n$, where $n$ is an integer, alongside an axial pitch vector along the tube axis, $q_z$ \cite{SuppMater}. In a cylindrical coordinate system, the magnetization can thus be modeled as 
\begin{align}
    \mathbf{M} = M_{\rm s}\begin{pmatrix}
    \cos(n\phi+q_{z} z + \psi)\sin\Theta\\  \sin(n\phi+q_{z} z + \psi)\sin\Theta \\
    \cos\Theta
    \end{pmatrix},
\label{eq:conical-helix-tube}
\end{align}
where $\psi$ denotes a phase arising from the competition between dipolar interactions and magnetic radial anisotropy, and $\Theta$ is the cone angle with respect to the tube axis. In this model, the curling state corresponds to the particular case $n=0$, $q_z=0$, and $\psi=\pi/2$. The volume and surface toroidal moments associated with these textures are only nonzero for the curling state with $\boldsymbol{\tau}^{\rm{v,s}}\parallel \hat{z}$ and the uniform transverse mode ($n=-1$) with finite $q_z$ and $\boldsymbol{\tau}^{\rm{v,s}}\perp \hat{z} $ as illustrated in Fig.~\ref{fig:tubes_dmi_textures}. Nonetheless, the surface toroidal moment ($\boldsymbol{\tau}^{\rm s}$) is also nonzero for the $n=1$ state \cite{SuppMater} something that Eq.~\eqref{Eq4} does not capture since $\boldsymbol{\tau}^{\rm v}=0$. 
Thus, based on the toroidal moment calculations, only the states $n=0,\pm 1$ could exhibit frequency nonreciprocity of dipolar origin, whereas states such as the hedgehog configuration ($n=0$, $q_z=0$ and $\psi=0$) or other do not support any toroidal moment and therefore do not produce asymmetry in the spin-wave dispersions.

\begin{figure}[!ht]
    \includegraphics[width=8.6cm]{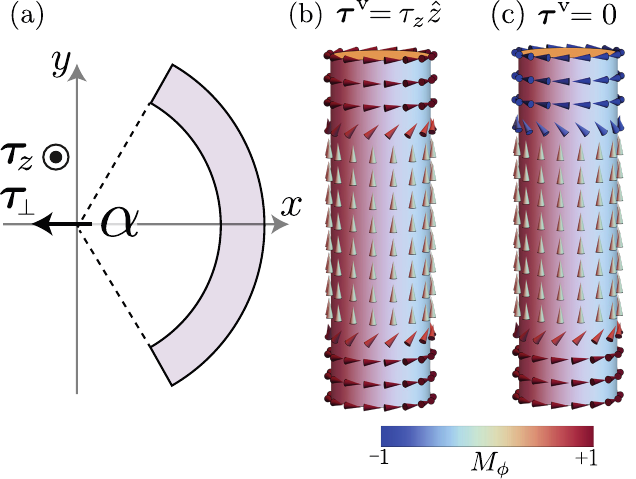}
	\caption{(a) Transversal cross-section of an incomplete tube with both axial and azimuthal magnetization components. Angle $\alpha$ defines the tubular section, such that $\alpha=2\pi$ stands for a fully closed tube. As illustrated, the volume toroidal moment possesses both axial ($\tau_z$) and transverse ($\tau_{\perp}$) components relative to the nanotube axis. (b-c) illustrate two closed nanotubes in an axial magnetization state, each with symmetric vortex domains at the end caps. In (b), the end vortices have the same helicity, producing a finite toroidal moment aligned with the tube axis. In (c), the vortices present opposite helicities, resulting in a zero toroidal moment.}
	\label{fig:tube_section_vortex}
\end{figure}

A more general magnetization configuration for thin cylindrical shells can be considered here, which can be written as $\mathbf{M}=M_{\phi}(z)\hat{\phi}+M_{z}(z)\hat{z}$. This expression can model border states at the tube caps \cite{Landeros09,Landeros22}, multivortex states \cite{Gan14,Streubel16}, and curved magnetic shells for partially closed tubes \cite{Brevis24}. Such a generalized model may produce toroidal moments along different directions depending on the symmetries and explicit dependencies. For instance, in partially closed tubes up to an angle $\alpha\leq 2\pi$, as illustrated in Fig. \ref{fig:tube_section_vortex}(a), in-plane of $\boldsymbol{\tau}$ components may arise depending on the value of $\alpha$ and the symmetry properties of $M_{\phi}(z)$. The toroidal moment for this configuration is given by $\boldsymbol{\tau}^{\rm v} =  
    \tau_{\perp} \hat{x}+ \tau_z \hat{z}$,
where 
\begin{equation}
    \tau_{\perp} =  -\frac{R
^2-R_i^2}{2}\sin(\frac{\alpha}{2})\int_{-L/2}^{L/2}{dz\,z\,M_\phi(z)}
\end{equation}
and 
\begin{equation}
\tau_z = \frac{R^3-R_i^3}{6}\left[\alpha +\frac{4}{\alpha} \sin^2\left(\frac{\alpha}{2}\right) \right]\int_{-L/2}^{L/2}{dz\,M_\phi(z)}.
\end{equation}
It can be seen that $\tau_{\perp}\neq 0$ for antisymmetric profiles of $M_\phi(z)$ with respect to the center of the tube. In the limit $\alpha=2\pi$, the fully closed tube is recovered, and only the $z-$component of the toroidal moment remains, but determined by $\int dz\, M_{\phi}(z)$. From this, it can be easily seen, for example, that a shell with an axial state in bulk but with vortex domain states at the caps, as is illustrated in Fig. \ref{fig:tube_section_vortex}(b,c), present a finite toroidal moment only if both vortex states have the same helicity. Otherwise, the toroidal moment cancels out. In summary, for partially closed tubes, transversal components of $\boldsymbol{\tau^{\rm v}}$ arise if $M_\phi (z)$ is asymmetric. For closed tubes, there are no transversal components. In both cases, the axial component is non-zero for symmetric $M_\phi (z)$ profiles. A similar result has been reported in Ref.~\citenum{Brevis24} where the propagation of spin waves in waveguides is studied as they are curved from flat to a tubular cross-section through intermediate states, which are equivalent to a system with an angle $\alpha < 2\pi$. The flat waveguide is always reciprocal, whose equilibrium magnetization yields zero toroidal moment. While all other configurations, as long as it has an equilibrium magnetization with azimuthal components producing toroidal moment parallel to the axis, are all nonreciprocal.

\subsection{Magnon nonreciprocity in graded films, bilayers and multilayers}
\label{subsec:graded_multilayers}

\begin{figure}[!ht]
	\includegraphics[width=8.8cm]{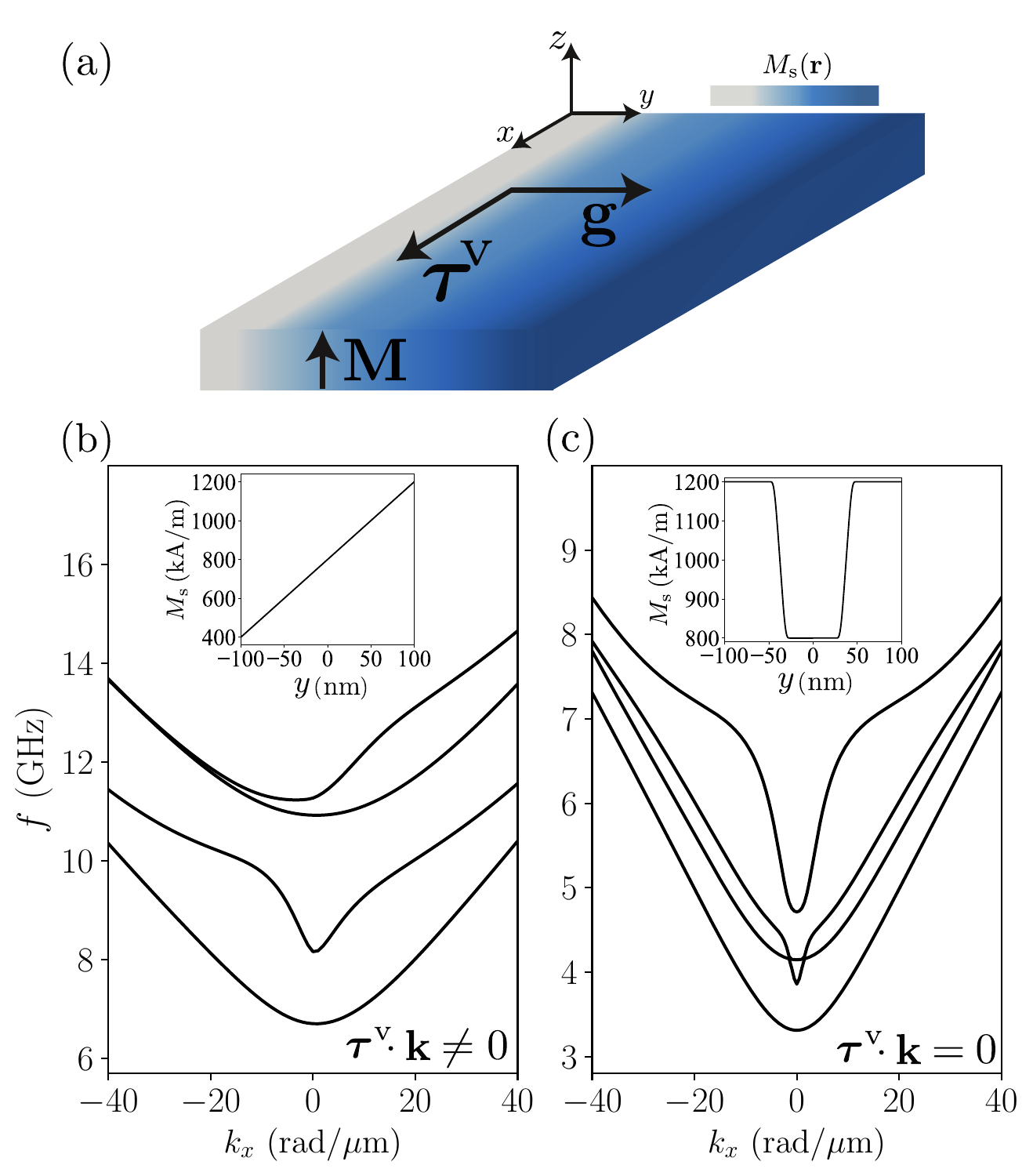}
	\caption{(a) Graded ferromagnetic stripe with $\mathbf{M}=M_{\rm{s}}(x_{\rm{g}})\hat{z}$, where the saturation magnetization changes along a specific direction $\mathbf{g}=\hat{y}$. According to Eq.~\eqref{tg}, this graded system exhibits a toroidal moment always oriented perpendicular to $\mathbf{M}$ and $\mathbf{g}$, with a magnitude that depends on the gradient profile. (b–c) Spin-wave dispersion spectra obtained from \textsc{TetraX} for stripes with magnetization gradients, showing the four lowest-frequency modes. In (b), the saturation magnetization profile is linear and asymmetric, resulting in a toroidal moment and nonreciprocal spin-wave propagation. In (c), the magnetization profile is mirror-symmetric concerning the film center, leading to a vanishing toroidal moment and a reciprocal dispersion. Insets show the respective magnetization saturation profiles across direction $x$.}
	\label{fig:graded_film}
\end{figure}

From Eq.~\eqref{Eq4}, one can note that a uniform $\mathbf{M}$ results in $\boldsymbol{\tau}^{\rm v}=0$. However, if only $M_{\rm{s}}$ is spatially dependent, a nonzero $\boldsymbol{\tau}^{\rm v}$ can arise even if a field polarizes the magnetization. Consider $\mathbf{M} = M_{\rm{s}} (\mathbf{r}) \mathbf{m}$, with $\mathbf{m} = (m_x,m_y,m_z)$ being the unitary magnetization. If the graduation of $M_{\rm{s}}$ along an arbitrary direction ($\mathbf{g}$), i.e., $M_{\rm{s}} (\mathbf{r}) = M_{\rm{s}} (x_{g}) $, where $x_{\rm{g}}$ represents the graduation coordinate, then the average magnetization is $\langle\mathbf{M}\rangle= (1/L_g)\int{dx_{g}\,M_{s}(x_{g})}\,\mathbf{m} \equiv \langle M_{\rm s} \rangle \mathbf{m}$.
Therefore, the toroidal moment of the graduated film is
\begin{equation}
	\boldsymbol{\tau}^{\rm v}= \frac{V}{2 L_g}  (\mathbf{g}\times\mathbf{m}) \int_{-L_{g}/2}^{L_{g}/2}{x_{\rm{g}}\,[M_{\rm{s}} (x_{\rm{g}}) - \langle M_{\rm s} \rangle] dx_{\rm{g}}} .
	\label{tg}
\end{equation}
Clearly, if $M_{\rm{s}} (x_{g})$ is constant then the integral vanishes and $\boldsymbol{\tau}^{\rm v}=0$.
For a linear graduation profile  $M_{\rm{s}} (x_{\rm{g}}) = \langle M_{\rm{s}}\rangle + \frac{\Delta M_s}{L_{\rm{g}}}x_{\rm{g}}$, where $\Delta M_s$ represents the total variation of the saturation magnetization from one extreme to the other, the resulting toroidal moment reads
\begin{equation}
\boldsymbol{\tau}^{\rm v} =\frac{V L_{\rm{g}}}{24} \Delta M_{\rm{s}}\,(\mathbf{g}\times\mathbf{m}).
\label{Tgraded}
\end{equation}
An example of a graduated system is depicted in Fig.~\ref{fig:graded_film}(a). Based on this, it follows from Eq.~\eqref{tg} that if  $(\mathbf{g} \times \mathbf{m}) \parallel \mathbf{k}$, then SW nonreciprocity would occur as predicted for Damon-Eshbach modes in a saturated film graded along the normal with different asymmetric profiles respect to the center \cite{Gallardo19a}. 
In contrast, the system should be reciprocal if $(\mathbf{g}\times\mathbf{m}) \perp \mathbf{k}$, as reported in \cite{Gallardo2022b} for the particular case of a laterally graded stripe. Note that the integral vanishes for any symmetric profile centered at $x_{\rm{g}}=0$. Therefore, Eq.~\eqref{tg} suggests that for graded films $\Delta f \propto (\mathbf{g}\times\mathbf{m})\cdot \mathbf{k}$, which generalizes the results in Refs. \onlinecite{Gallardo19a,Gallardo2022b} for small wave vectors, and allows to predict the existence of nonreciprocity in a more general way.
This can be seen in Fig.~\ref{fig:graded_film}, where a thin magnonic waveguide magnetized in the normal direction ($z$) is graduated along the width ($y$), and the spin waves propagate along $x$. In Fig.~\ref{fig:graded_film}(c), a linear (asymmetric) graded magnetization profile is considered, resulting in a nonreciprocal dispersion, while in Fig.~\ref{fig:graded_film}(d), a symmetric profile results in  $\Delta f=0$, since $\boldsymbol{\tau}^{\rm v}=0$, regardless of the orientation of the wave vector. Both dispersions were obtained using the \textsc{TetraX}  micromagnetic package \cite{TetraX,Korber21c}. The simulated system is a stripe with a cross-sectional width of 200~nm and a thickness of 60~nm, assuming translational invariance along the spin-wave propagation direction ($x$-axis). The magnetic parameters used are a gyromagnetic ratio $\gamma = 185.66$ GHz/T and an exchange stiffness constant $A_{\rm{ex}}=10$~pJ/m. To saturate the sample, an external magnetic field of $B_{\rm{ext}} = 1$~T was applied along the normal direction. Two distinct saturation magnetization profiles were considered across the stripe width: a linear gradient and a symmetric profile. In both cases, the saturation magnetization $M_{\rm s}$ ranges from a minimum of 400~kA/m to a maximum of 1200~kA/m [see insets of the Figure~\ref{fig:graded_film}(c-d)].
A specific class of magnetization-graded systems can be achieved using two ferromagnetic layers made of different materials in direct contact, each one uniformly magnetized. Here, $M_{\rm{s}}(x_{\rm{g}})$ changes abruptly, and the bilayer can then be regarded as a graded system with a sharp step in $M_{\rm{s}}$ at the interface. Nonreciprocal waves have been measured with Brillouin light scattering (in Damon-Eshbach configuration) in parallelly magnetized Ni/NiFe \cite{Mruczkiewicz17}, and CoFeB/NiFe bilayers \cite{Grassi20}, in perfect agreement with the predictions of Eq.~\eqref{tg}. In a thick film that has a graduation of the saturation magnetization along the normal direction, it has been shown that $\Delta f(k)$ increases linearly, reaches a maximum value, and then decreases \cite{Gallardo19a}. Because there is no analytical formula describing the associated magnetic tensor $\mathcal{N}^{(21)}(\mathbf{k})$ nor the frequency shift, the toroidal moment evaluation constitutes a very useful tool to anticipate the existence of nonreciprocity in graded magnetic films, which has only been determined numerically \cite{Gallardo19a} and more recently measured in N-doped Fe films \cite{Christienne25}, in full agreement with the insights provided by the toroidal moment in Eq.~\eqref{tg}.

\begin{figure}[!ht]
	\includegraphics[width=8.8cm]{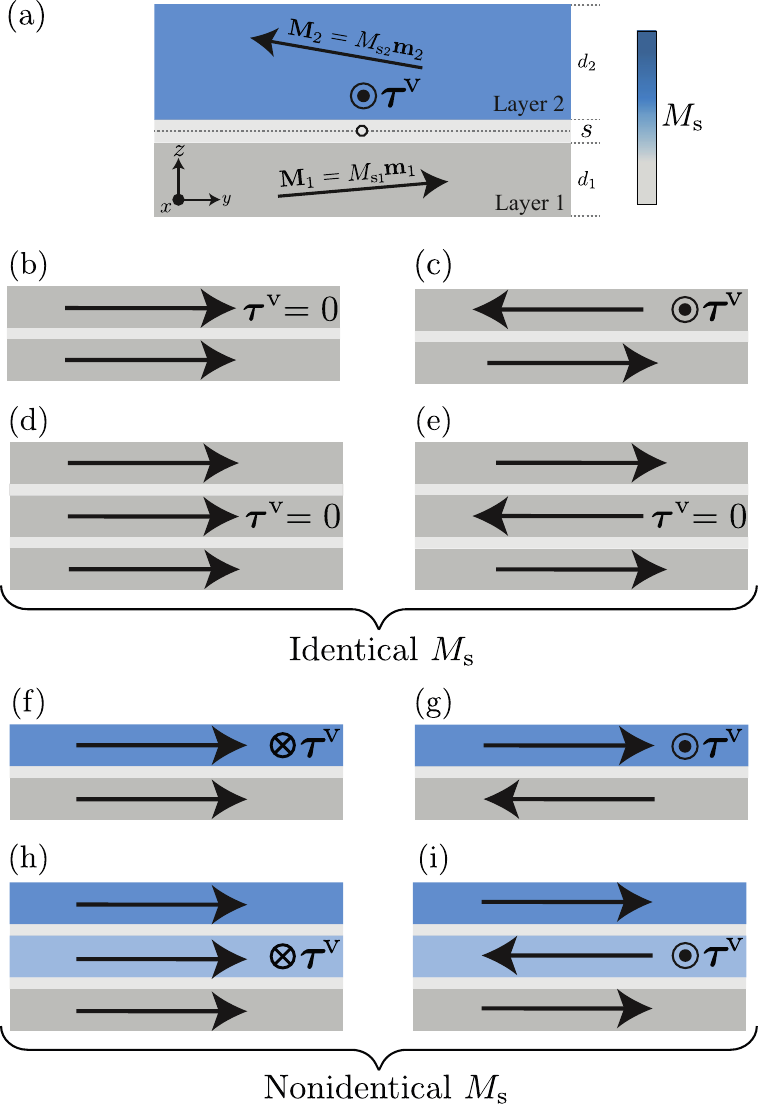}
	\caption{(a) Asymmetric ferromagnetic bilayer having thicknesses $d_{1,2}$ and magnetizations $\mathbf{M}_{1,2} = M_{\rm{s}1,2} \mathbf{m}_{1,2}$ separated by an spacer $s$. (b-e) illustrate bilayer (b-c) and trilayer (d-e) configurations with parallel and antiparallel magnetization orientations (indicated by black horizontal arrows), assuming identical $M_{\rm s}$. Here, only the bilayer with an antiparallel configuration has a toroidal moment. The other cases have $\boldsymbol{\tau}^{\rm v}=0$. (f-i) show similar configurations but with nonidentical $M_{\rm s}$ across layers (see color code), showing a finite toroidal moment for both parallel and antiparallel arrangements. In all cases (b–i), the layers have equal thickness.}
	\label{fig:bilayer_trilayer}
\end{figure}

In the specific case of a bilayer composed of two magnetic films with top/bottom surface area $L^2$, different thicknesses ($d_{1,2}$) and saturation magnetizations ($M_{\rm{s}1,2}$), separated by a nonmagnetic spacer of thickness $s$ [see Fig.~\ref{fig:bilayer_trilayer}(a)], depending on whether the magnetization in the lower layer $\mathbf{M}_1 = M_\mathrm{s1}\mathbf{m}$ and upper layer $\mathbf{M}_2 = \sigma M_\mathrm{s2}\mathbf{m}$ are parallel ($\sigma =+1$) or antiparallel ($\sigma=-1$), the average magnetization field is
\begin{align}
    \langle \mathbf{M} \rangle = \frac{M_{\rm{s}1}d_1 + \sigma M_{\rm{s}2} d_2 }{d_1 + d_2} \mathbf{m} \equiv \langle M \rangle \mathbf{m},
\end{align}
and the toroidal moment, according to Eq.~\eqref{tg}, becomes
\begin{equation}
    \boldsymbol{\tau}^{\rm v} = -\frac{L^2}{4} \frac{d_1 d_2 (d_1+d_2+2 s)}{(d_1+d_2)} (M_{\rm{s1}}-\sigma M_{\rm{s2}})\, (\hat{z} \times \mathbf{m}).
\label{eq:tv_bilayer}
\end{equation}
This formula explicitly shows that the toroidal moment vanishes only for bilayers with parallel magnetizations ($\sigma = 1$) and $M_{\rm{s}1} = M_{\rm{s}2}$, as is schematized in Fig.~\ref{fig:bilayer_trilayer}(b). In all other cases, where the bilayers have antiparallel magnetizations, or they are parallel but with different $M_{\rm{s}}$, the toroidal moment calculation suggests that nonreciprocity occurs for waves propagating parallel to $(\hat{z} \times \mathbf{m})$, with $z$ being normal to the bilayer surface \cite{Gallardo19b,Sluka19,Albisetti19}, as shown in Fig.~\ref{fig:bilayer_trilayer}(c,f,g).  
Nevertheless, these calculations can be extended to predict nonreciprocity in much more complex systems, such as multilayers that can be composed by different magnetic materials and thicknesses (calculations not shown). 
Fig.~\ref{fig:bilayer_trilayer}(b-i) summarizes the toroidal  moments obtained for parallel (left) and antiparallelly (right) magnetized bilayers and trilayers, composed of  (b-e) identical and (f-i) nonidentical magnetic materials.  
Through simple calculations, it can be shown that an antiferromagnetically coupled multilayer system (with magnetizations oriented along $\pm\, y $), with identical $M_{\rm{s}}$ and thicknesses, exhibits a nonzero toroidal moment along the in-plane $x$-direction if the number of layers is even. 
In contrast, when the number of identical layers is odd, the toroidal moment is zero, as can be seen in Fig.~\ref{fig:bilayer_trilayer}(d-e) for a trilayer configuration. However, for any number of layers, a finite toroidal moment emerges if the symmetry is broken along the normal concerning the multilayer center. This symmetry breaking can be achieved through the modulation of $M_{\rm s}$, as depicted in Fig.~\ref{fig:bilayer_trilayer}(f-i), or thicknesses. This behavior applies to both ferromagnetic and antiferromagnetic coupled layers. 

For a symmetric bilayer ($d_1 = d_2 = d$) with two identical ferromagnetic layers but antiparallel ($\sigma=-1$) the toroidal moment from Eq.~\eqref{eq:tv_bilayer} is $ \boldsymbol{\tau}^{\rm v} = -\frac{L^2}{2} d(d+s)(\hat{z} \times \mathbf{M})$, such as depicted in Fig.~\ref{fig:bilayer_trilayer}(c). The antiparallelly magnetized symmetric bilayer is a particular case that allows for a simple analytical formula for the magnon frequency shift induced by the dynamic dipolar coupling in an extended wave vector regime, which leads to a magnetic tensor \cite{Gallardo19b} 
\begin{equation}
    {\rm Im}[\mathcal{N}^{(21)}(\mathbf{k})]=2\sinh^2\left({\frac{kd}{2}}\right)\frac{e^{-\mid k\mid(d+s)}}{kd},
\label{Eq-N21bilayer}
\end{equation}
which is nonlinear with $k$.
Only in the low-$k$ regime, ${\rm Im}[\mathcal{N}^{(21)}(\mathbf{k})]\approx kd/2$ and the frequency shift is linear with $k$, which allows to write $\Delta f \propto \boldsymbol{\tau} \cdot  \mathbf{k}\propto (\hat{z} \times \mathbf{M})\cdot  \mathbf{k}$.

Interestingly, the work of Kuznetsov and Fraerman \cite{Kuznetsov2022} theoretically investigated spin-wave nonreciprocity in hybrid films consisting of a thin ferromagnet (FM) coupled with either a paramagnet (PM) or a superconductor (SC) semi-infinite layer. In both cases, the nonreciprocity originates from the dipolar interaction between the FM layer and the PM/SC material, which breaks the symmetry along the normal direction. In the FM/PM case, the spin waves induce a dynamic magnetization in the paramagnet, which in turn creates an asymmetric dipolar field that influences the SW propagation. Then, the FM/PM bilayer behaves as the graded system presented earlier, where the induced dynamic magnetization in the paramagnet results in a symmetry breaking associated with a toroidal moment perpendicular to the equilibrium magnetization and the interface normal, in agreement with Eq.~(\ref{tg}). For the FM/SC system, SWs in the FM induce a superconducting current in the SC, which also creates an asymmetric dipole field that breaks the symmetry. This inhomogeneous field within the ferromagnet can be associated with a toroidal moment parallel to $ \hat{n}\times\mathbf{M}$ \footnote{In Ref. \cite{Kuznetsov2022}, the calculated frequency asymmetry $\Delta f$ is proportional to $\mathbf{k}\cdot ( \hat{n}\times\mathbf{M})$, which agrees with the nonreciprocity condition if $\boldsymbol{\tau} \rightarrow \hat{n}\times\mathbf{M}$.}.

\subsection{Magnon nonreciprocity in DMI films}

\begin{figure}[t]
	\includegraphics[width=8.6cm]{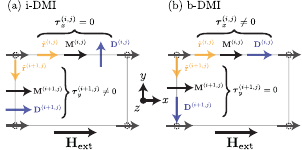}
	\caption{Bond magnetic toroidal moment for a square crystal lattice with magnetization and applied field parallel to the $x$-axis for a thin film with (a) interfacial DMI and (b) bulk DMI. The gray arrows at the corners represent the magnetic moment of each site (all oriented in the $x$-direction), while the yellow arrows depict the bond unit vectors $\hat{\mathbf{r}}^{(i,j)}$ and $\hat{\mathbf{r}}^{(i+1,j)}$ along $x$ and $y$ directions. Then, with the bond vector, the average magnetization, and the DM vector for each pair, the BMTD associated with each configuration can be obtained. 
 }
	\label{fig:fig8}
\end{figure}

Dzyaloshinskii-Moriya materials may also present a toroidal moment \cite{Tan24}. For interfacial DMI, nonreciprocity is present in flat films for $\mathbf{M}\perp \mathbf{k}$, meanwhile for bulk DMI, nonreciprocity appears for $\mathbf{M}\, ||\, \mathbf{k}$ \cite{Cortes13,Gallardo19BCh,Kuepferling23}.  
From microscopic calculations, modulations in the dispersion bands can be expressed through magnetic toroidal bond multipoles linked to anisotropic current distributions in some systems \cite{Hayami20a,Hayami20b}. These expressions are useful for predicting or estimating nonreciprocity and transport properties. From such a multipole description for microscopic systems, it is also possible to obtain a toroidal moment associated with the type of DMI, the crystal lattice, and the magnetic configuration \cite{Matsumoto21}. This quantity, referred to as bond magnetic toroidal dipole (BMTD), is a microscopic indicator of DMI-induced nonreciprocity and fulfills the condition $\boldsymbol{\tau} \cdot \mathbf{k} \neq 0$ for asymmetric dispersions. The BMTD is calculated using $\boldsymbol{\tau}^{(ij)}\cdot \hat{\mathbf{r}}^{(ij)} = \mathbf{D}^{(ij)} \cdot \mathbf{M}^{(ij)} $, where $\mathbf{M}^{(ij)}$ is the averaged magnetic moment for the $i$th and $j$th spins, $\hat{\mathbf{r}}^{(ij)}$ is the unit vector connecting these spins, and $\boldsymbol{\tau}^{(ij)}$ represents the bond magnetic toroidal dipole on the bond ($ij$)  \cite{Matsumoto21}. Therefore, the BMTD is active if $\mathbf{D}^{(ij)}\cdot\mathbf{M}^{(ij)}\neq 0$ and then, nonreciprocity due to DMI arises for magnons moving parallel to BMTD. For interfacial DMI, it is known that the DM vector ($\mathbf{D}^{(ij)}$) is perpendicular to $\hat{\mathbf{r}}^{(ij)}$ and nonreciprocity appears for $\mathbf{M}\perp \mathbf{k}$, meanwhile for bulk DMI the DM vector is parallel to $\hat{\mathbf{r}}^{(ij)}$, and nonreciprocity can occur for $\mathbf{M}\, ||\, \mathbf{k}$, in perfect agreement with well-known results \cite{Cortes13,Gallardo19BCh,Kuepferling23}. These two situations are illustrated in Fig.~\ref{fig:fig8}, where two square lattices are schematized at whose sites there are localized spins in the presence of DMI. These magnetic moments are aligned along the $x$-axis due to a bias field so that the average magnetization in each bond points in that direction. In Fig.~\ref{fig:fig8}(a), interfacial DMI is considered, while in Fig.~\ref{fig:fig8}(b), bulk DMI is depicted. Thus, in (a) [(b)] the DM vector points perpendicular [parallel] to the vector joining two consecutive sites ($\hat{\mathbf{r}}^{(i,j)}$), representing the condition for an active BMTD to occur on each bond.

\section{Final Remarks}
To conclude, the relationship between different definitions of the toroidal moment applied to confined magnetic structures has been elucidated, where a surface toroidal moment arising from surface-bound currents connects them without the need for time-averaging. By comparing these quantities applied to well-known magnetic textures, it is found that they are always parallel and may differ in magnitude and sign. For this reason, the surface toroidal moment can be very useful from the experimental point of view for samples with homogeneous textures: it is enough to know the surface magnetization to estimate its toroidal moment. Since one of the main motivations in this work relates to the prediction of magnon nonreciprocity, which is primarily determined by the toroidal moment direction and not by its magnitude, the discrepancy in the magnitude of the quantities is irrelevant. The dipolar nonreciprocity is discussed when the propagation direction of spin waves has a component parallel to the toroidal moment.
Furthermore, a criterion is established for calculating it, thereby avoiding the issue of selecting the correct origin based on compensated and uncompensated systems. Finally, this criterion is applied to various nonreciprocal systems, where the calculations are consistent with those reported in the literature and predict the existence of nonreciprocity in a more general way. Additionally, a toroidal moment for systems with DMI is discussed, demonstrating nonreciprocity under given conditions. These results could expand the discussion of the toroidal moment and facilitate the identification and estimation of nonreciprocity in magnonic systems mediated by dipolar and antisymmetric exchange interactions.

\begin{acknowledgments}
\textit{Acknowledgments}--- 
We are thankful to Prof. Paula Mellado and Prof. Phillip Pirro for fruitful discussions.
F.B. acknowledges support from ANID National Doctoral scholarship 2021-21211469 and from USM through PIIC program. Financial support from Fondecyt grants 1241589 and 1250803, and Basal Program for Centers of Excellence, Grant AFB220001 (CEDENNA), is kindly acknowledged. L.K. gratefully acknowledges funding by the Radboud Excellence Initiative.
\end{acknowledgments}


\providecommand{\noopsort}[1]{}\providecommand{\singleletter}[1]{#1}%

\end{document}


\title{Supplementary material for\\
\textit{Toroidal Moments in Confined Nanomagnets and their Impact on Magnonics}}

\author{F. Brevis}
\email{felipe.brevis@usm.cl}
\affiliation{Departamento de Física, Universidad Técnica Federico Santa María, Avenida España 1680, Valparaíso,
Chile}
\author{L. K\"orber}
\affiliation{Radboud University, Institute of Molecules and Materials, Heyendaalseweg 135, 6525 AJ Nijmegen, The Netherlands}
\author{B. Mimica-Figari}
\affiliation{Departamento de Física, Universidad Técnica Federico Santa María, Avenida España 1680, Valparaíso,
Chile}
\author{R. A. Gallardo}
\affiliation{Departamento de Física, Universidad Técnica Federico Santa María, Avenida España 1680, Valparaíso,
Chile}
\author{A. K\'akay}
\affiliation{Helmholtz-Zentrum Dresden-Rossendorf, Institute of Ion Beam Physics and Materials Research, Bautzner Landstr. 400, 01328 Dresden, Germany}
\author{P. Landeros}
\email{pedro.landeros@usm.cl}
\affiliation{Departamento de Física, Universidad Técnica Federico Santa María, Avenida España 1680, Valparaíso,
Chile}

\date{\today}
\begin{abstract}
  This Supplementary Material provides a detailed account of the main calculations, starting with the derivation of the surface toroidal moment. Next, the expressions for the toroidal moment arising from well-known magnetic textures are computed. 

\end{abstract}
\maketitle
\section{Analytical derivation of the surface toroidal moment}
The toroidal moment associated with any current proposed by Dubovik and Tugushev \cite{Dubovik90} is
\begin{align}
    \boldsymbol{\tau}=\frac{1}{10}\int dV\, \left[\mathbf{r}(\mathbf{r}\cdot \mathbf{J})-2r^2 \mathbf{J}\right].
\label{eq:dubovik}
\end{align} 
In order to investigate the connection between the magnetization ($\mathbf{M}$) and the toroidal moment ($\boldsymbol{\tau}$), consider that a magnetization distribution may create a bound current $(\mathbf{J_b})$, defined as
\begin{align}
    \mathbf{J_b}(\mathbf{r})=\nabla \times \mathbf{M},
\end{align} 
and in the absence of free and polarization currents, $\mathbf{J} = \mathbf{J}_{\rm{b}}$. Replacing this expression into the Eq (\ref{eq:dubovik}),
\begin{align}
    \boldsymbol{\tau}=\frac{1}{10}\int dV\, \left[\mathbf{r}(\mathbf{r}\cdot (\nabla \times \mathbf{M}))-2r^2 (\nabla \times \mathbf{M})\right].
\label{eq:dubovikcurlM}
\end{align}
As $\nabla \times \mathbf{r}=0$, then, the divergence of a cross-product can be written as
\begin{align}
    \nabla \cdot (\mathbf{r}\times\mathbf{M})&=-\mathbf{r}\cdot(\nabla\times\mathbf{M}).
\end{align}
On the other hand, the curl of the product between a scalar function $\psi$ and a vector function $\mathbf{M}$ can be expanded as
\begin{align}
    \nabla \times(\psi \mathbf{M})&=\psi(\nabla \times \mathbf{M})+(\nabla\psi)\times\mathbf{M}.
\end{align}
Considering that $\psi=-2r^2$, the second term in Eq.~\ref{eq:dubovikcurlM} gives 
\begin{align}
    -2r^2(\nabla \times \mathbf{M}) =  -2\nabla \times(r^2 \mathbf{M})+4\mathbf{r}\times\mathbf{M},
\end{align}
then, the toroidal moment becomes
\begin{align}
\boldsymbol{\tau} = \frac{1}{10}\int dV\, \left[\ -\mathbf{r}[\nabla \cdot (\mathbf{r}\times\mathbf{M})] -2\nabla \times(r^2 \mathbf{M})+4\mathbf{r}\times\mathbf{M}  \right].
\label{Eq.TM7}
\end{align}
From vector calculus identities, 
\begin{align}
    \int_{V}{dV\, \psi \nabla\cdot\mathbf{A}} =\oiint_{S}{\psi \mathbf{A}\cdot d\mathbf{S}} - \int_{V}{dV\mathbf{A}\cdot\nabla\psi}.
\end{align}
On the other hand, in Cartesian coordinates, it can be considered that,
\begin{align}
    \int dV\, \mathbf{r}[\nabla \cdot (\mathbf{r}\times\mathbf{M})] = \hat{r}_i\int dV\, r_i[\nabla \cdot (\mathbf{r}\times\mathbf{M})] 
\end{align}
Where $r_i=x,y,z$, and the Einstein notation is used. Considering $\psi = r_i$ and $\mathbf{A} = \mathbf{r}\times\mathbf{M}$, then
\begin{align}
    \int_{V}{dV\, r_i \nabla\cdot(\mathbf{r}\times\mathbf{M})} =\oiint_{S}{r_i (\mathbf{r}\times\mathbf{M})\cdot d\mathbf{S}} - \int_{V}{dV\,(\mathbf{r}\times\mathbf{M})\cdot\nabla r_i},
\end{align}
but we have that $\nabla r_i = \hat{r}_i$, then
\begin{align}
    \int_{V}{dV\, r_i \nabla\cdot(\mathbf{r}\times\mathbf{M})} =\oiint_{S}{r_i (\mathbf{r}\times\mathbf{M})\cdot d\mathbf{S}} - \int_{V}{dV\,(\mathbf{r}\times\mathbf{M})_{i}},
\end{align}
and aadding the three components,
\begin{align}
    \hat{r}_i\int_{V}{dV\, r_i \nabla\cdot(\mathbf{r}\times\mathbf{M})} = \hat{r}_i\oiint_{S}{r_i (\mathbf{r}\times\mathbf{M})\cdot d\mathbf{S}} - \int_{V}{dV\,(\mathbf{r}\times\mathbf{M})}.
\end{align}
Replacing the obtained result into the toroidal moment in Eq.~\ref{Eq.TM7},
\begin{align}
    \boldsymbol{\tau} = -\frac{1}{10} \oiint_{S}{dS\,\mathbf{r}[ (\mathbf{M} \times \mathbf{n})\cdot \mathbf{r} }] - \frac{2}{5} \int{ dV\,\,\nabla \times(r^2 \mathbf{M})}+\frac{1}{2}\int{dV\,\,\mathbf{r}\times\mathbf{M}}.
\end{align}
The volume integral of the curl of a function, which can be obtained from the Gauss Theorem, is
\begin{align}
    \int_V{dV\,\nabla \times\mathbf{\mathcal{G}}} = -\oiint_{S}{\mathcal{G}\times d\mathbf{S}}.
\end{align}
By considering that $\mathcal{G}= r^2 \mathbf{M} $ and $d\mathbf{S}=\mathbf{n}\,dS$ where $\mathbf{n}$ is the unit vector normal to the surface,
\begin{align}
    \boldsymbol{\tau} =-\frac{1}{10}\oiint_{S}{dS\,\mathbf{r}[ (\mathbf{M} \times \mathbf{n})\cdot \mathbf{r} }]  + \frac{1}{5} \oiint_{S}{dS\,r^2 (\mathbf{M} \times \mathbf{n}) } +\frac{1}{2}\int{dV\,\,\mathbf{r}\times\mathbf{M}}.
\end{align}
 It follows that $\mathbf{M} \times \mathbf{n}$ corresponds to the bound surface current $\mathbf{K}_b$ \cite{griffiths2013introduction}, and the toroidal moment is
\begin{align}
    \boldsymbol{\tau} = \frac{1}{2}\int_{V}{dV\,\,(\mathbf{r}\times\mathbf{M})} - \frac{1}{10}\oiint_{S}{dS\,\left[\mathbf{r}(\mathbf{r}\cdot \mathbf{K}_b)- 2r^2 \mathbf{K}_b\right]}.
\end{align}
Then, the toroidal moment splits into two contributions, $\boldsymbol{\tau} = \boldsymbol{\tau}^{\rm{v}} + \boldsymbol{\tau}^{\rm{s}}$, where
\begin{align}
	\boldsymbol{\tau}^{\rm{v}} = \frac{1}{2}\int_{V}{dV\,(\mathbf{r}\times\mathbf{M})}
    \label{eq:tau_v}
\end{align}
is the volume term and the standard expression to evaluate the toroidal moment \cite{Spaldin08, Bhowal2022}, and 
\begin{align}
	\boldsymbol{\tau}^{\rm{s}} = - \frac{1}{10}\oiint_{S}{dS\,\left[\mathbf{r}(\mathbf{r}\cdot \mathbf{K}_b)- 2r^2 \mathbf{K}_b\right]}
    \label{eq:tau_s}
\end{align}
is the emergent toroidal moment due to surface-bound current $\mathbf{K}_{\rm{b}}$. Notice that $\boldsymbol{\tau}^{\rm{s}}$ is analogous to Eq.~(\ref{eq:dubovik}) but with a closed surface integral and opposite sign.
\section{Toroidal moments of the studied magnetic textures}
In this section, the volume and surface toroidal moments are calculated for various noncollinear magnetic states including the conical-helix textures in planar and tubular systems and skyrmionic magnetizations such as the skyrmion, bimeron and meron.

\subsection{Analytical expressions for toroidal moments from conical-helix texture.}
Using the model for the conical-helix texture magnetization described in the main text, the surface and volume toroidal moments, $\boldsymbol{\tau}^{\rm{s}}$ and $\boldsymbol{\tau}^{\rm{v}}$ respectively, calculated with respect to the geometric center of an ultrathin squared film with area $L^2$ and thickness $d$, are given by
\begin{align}
   \boldsymbol{\tau}^{\rm{s}}= \sin\theta
    \begin{pmatrix} 
        \mathcal{A}^{\rm{s}}_{1}\sin\psi\cos\varphi_{\mathbf{q}} \\ 
       -\mathcal{A}^{\rm{s}}_{1}\sin\psi \sin\varphi_{\mathbf{q}} \\ 
        \mathcal{A}^{\rm{s}}_{2}\cos\psi \cos\varphi_{\mathbf{q}}
    \end{pmatrix}    
\label{eq:ch_toroidal_moment_s}
\end{align}
and
\begin{align}
   \boldsymbol{\tau}^{\rm{v}}= \mathcal{B}^{\rm{v}}\sin\theta
    \begin{pmatrix} 
        \sin\psi\cos\varphi_{\mathbf{q}} \\ 
       -\sin\psi \sin\varphi_{\mathbf{q}} \\ 
        \cos\psi \cos\varphi_{\mathbf{q}}
    \end{pmatrix},
\label{eq:ch_toroidal_moment_v}
\end{align}
where $\mathcal{A}^{\rm{s}}_{1},\mathcal{A}^{\rm{s}}_{2}$ and $\mathcal{B}^{\rm{v}}$ are
\begin{align}
	\mathcal{A}^{\rm{s}}_{1} = \frac{M_{\rm{s}} d L }{60 q^2}\left[\left(2 d^2 q^2+7 L^2 q^2+12\right)\sin \left(\frac{L q}{2}\right) -6 L q \cos \left(\frac{Lq}{2}\right)\right],
\end{align}
\begin{align}
	\mathcal{A}^{\rm{s}}_{2} =  \frac{M_{\rm{s}} d L }{60 q^2}\left[\left(d^2 q^2+8 L^2 q^2+12\right) \sin \left(\frac{L q}{2}\right) -6 L q \cos \left(\frac{Lq}{2}\right)\right], 
\end{align}
\begin{align}
	\mathcal{B}^{\rm{v}} = \frac{M_{\rm{s}} d L}{2q^2}  \left[L q \cos \left(\frac{Lq}{2}\right)-2 \sin \left(\frac{Lq}{2}\right)\right].
\end{align}
An oscillatory behavior around $Lq/2$ is a prominent feature, which may result in positive, zero, or negative toroidal moments. In the ultrathin limit $L\gg d$, all the terms proportional to $(qd)^2$ are negligible. When comparing $\boldsymbol{\tau}^{{\rm{v}}}$ with $\boldsymbol{\tau}^{\rm{s}}$ it is interesting to note that despite their differences, the resulting toroidal moments are parallel to each other and always perpendicular to the respective pitch vector ($\mathbf{q}$) despite the choice of origin. Table \ref{sm:table1} summarizes the toroidal moment directions for this texture and for both interfacial and bulk Dzyaloshinskii-Moriya interactions (b-DMI and i-DMI).

\begin{table}[htb]
\fontsize{10pt}{10pt}
\caption{Summary of the surface toroidal moment directions for the conical helix with bulk and interfacial DMI and for the two relevant phase constants appropriate to in-plane ($\psi=\pi/2$) and out-of-plane ($\psi=0$) magnetic films. The brackets show the pairs of surfaces contributing to $\boldsymbol{\tau}^{\rm{s}}$.}
\renewcommand{\arraystretch}{1.25} 
\begin{ruledtabular}
\begin{tabular}{ccc}
\textbf{Conical-helix} & \textbf{b-DMI} & \textbf{i-DMI} \\
 $\psi$ &  $(\varphi_{\mathbf{q}}=0$, $\mathbf{q}=q\,\hat{y})$ & $(\varphi_{\mathbf{q}}=\pi/2$, $\mathbf{q}=q\,\hat{x})$ \vspace{0.5mm} \\
\colrule 
$\pi/2$ & $(\tau^{\rm s}_{x},0,0) | \{S_{x},S_{y}\}$ &  $(0,\tau^{\rm s}_{y},0) | \{S_{x},S_{y}\}$  \vspace{0.5mm} \\
$0$ & $(0,0,\tau^{\rm s}_{z}) | \{S_{y},S_{z}\}$ & $0$ \\
\end{tabular}
\end{ruledtabular}
\label{sm:table1}
\end{table}

\begin{figure*}
    \includegraphics[width=\columnwidth]{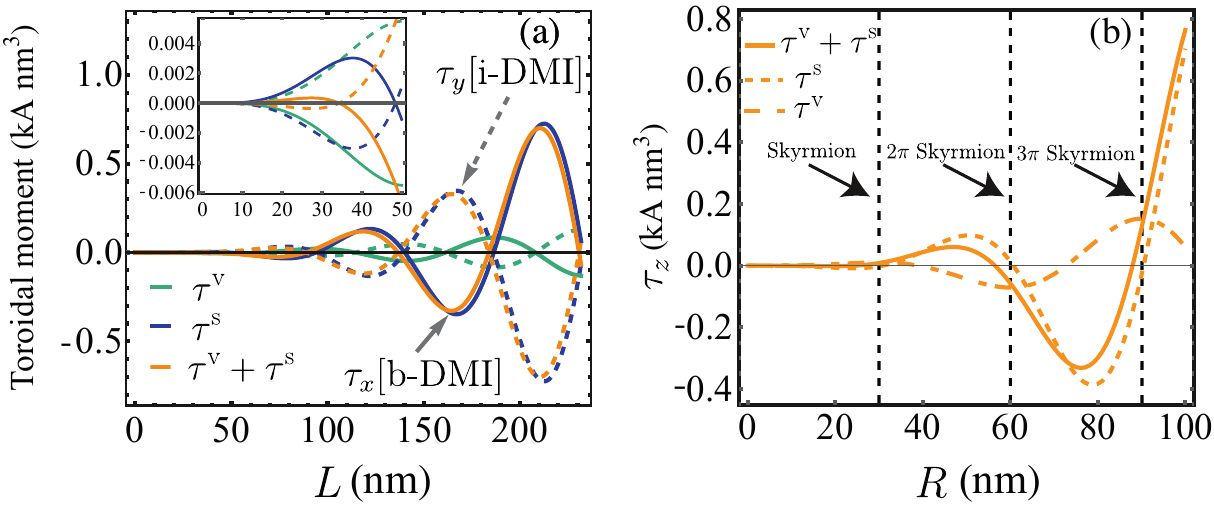}
    \caption{(a) Nonzero components of the toroidal moments $\boldsymbol{\tau}^{\rm{v}}$, $\boldsymbol{\tau}^{\rm{s}}$, and their sum $\boldsymbol{\tau}$, for the conical-helix texture as a function of the film size $L$. The DMI strength is fixed at $D = 3\,\rm{m}J/\rm{m}^2$, and the magnetic phase ($\psi=\pi/2$) correspond to a film with low perpendicular anisotropy. Continuous (segmented) curves correspond to bulk (interfacial) DMI with $x$($y$)-component, and the inset corresponds to a zoom of the first tens of nanometers. (b) Toroidal moments ($z-$component) for Bloch skyrmion-like states ($\gamma=\pi/2$) as a function of the nanodot radius when $\lambda=30$ nm. }
    \label{sm:fig1}
\end{figure*}

Fig.~\ref{sm:fig1} shows the dependence of $\boldsymbol{\tau}^{\rm{v}}$ with the lateral size $L$ for fixed parameters and for both types of DMI. It can be seen that a difference between b-DMI ($\varphi_\mathbf{q}=0$) and i-DMI ($\varphi_\mathbf{q}=\pi/2$) is the sign and the direction of $\boldsymbol{\tau}$ when $\psi=\pi/2$.  For the other case, $\psi=0$, a nonzero $\boldsymbol{\tau}$ is seen only for b-DMI, while for i-DMI, it cancels out. By analyzing the surface contribution, it is observed that for b-DMI, there is a slight difference in the amplitudes when comparing $\psi=0$ and $\psi=\pi/2$, which is not the case for $\boldsymbol{\tau}^{\rm{v}}$. 
In spite there is only one case in which $\boldsymbol{\tau}$ is always zero (for i-DMI and $\psi=0$), for all other configurations of $\varphi_{\mathbf{q}}$ and $\psi$, the toroidal moment could be zero for specific values of $Lq/2$, which is due to commensurability.  For the configurations studied, it is observed that $\boldsymbol{\tau}^{\rm{v}}$ and $\boldsymbol{\tau}^{\rm{s}}$ are always parallel to either $x$, $y$, or $z$ depending on the type of DMI ($\varphi_\mathbf{q}$) and phase ($\psi$).

In order to describe the surface toroidal moment ($\boldsymbol{\tau}^{\rm{s}}$) it is necessary to identify which faces contribute to its formation. Given the cuboid geometry of the sample, the surfaces are grouped by the direction of its normal vector, where $S_{x}$, $S_{y}$, and $S_{z}$ refers to each pair of surfaces. Table \ref{sm:table1} indicates the faces contributing to $\boldsymbol{\tau}^{\rm{s}}$ in curly brackets. It is found, surprisingly, that for low perpendicular anisotropy ($\psi=\pi/2$), the surfaces contributing to $\boldsymbol{\tau}^{\rm{s}}$ are the lateral edges ($S_{x}, S_{y}$) with area $Ld$. In contrast, for the surviving $\boldsymbol{\tau}$ when $\psi=0$, the contributing pair of surfaces are $S_{y}$ and $S_{z}$, where $S_{z}$ have an area $L^2$ much larger than for $S_{x,y}$. From the point of view of the amplitudes, no significant difference is predicted when comparing $\psi=0$ and $\psi=\pi/2$ for b-DMI as mentioned above; it is inferred then that the size of the areas is not particularly relevant, but rather the direction of their normal vectors.

\subsection{Analytical expressions for toroidal moments from Skyrmion, bimeron and meron.}

\begin{figure}[!ht]
	\includegraphics[scale=0.53]{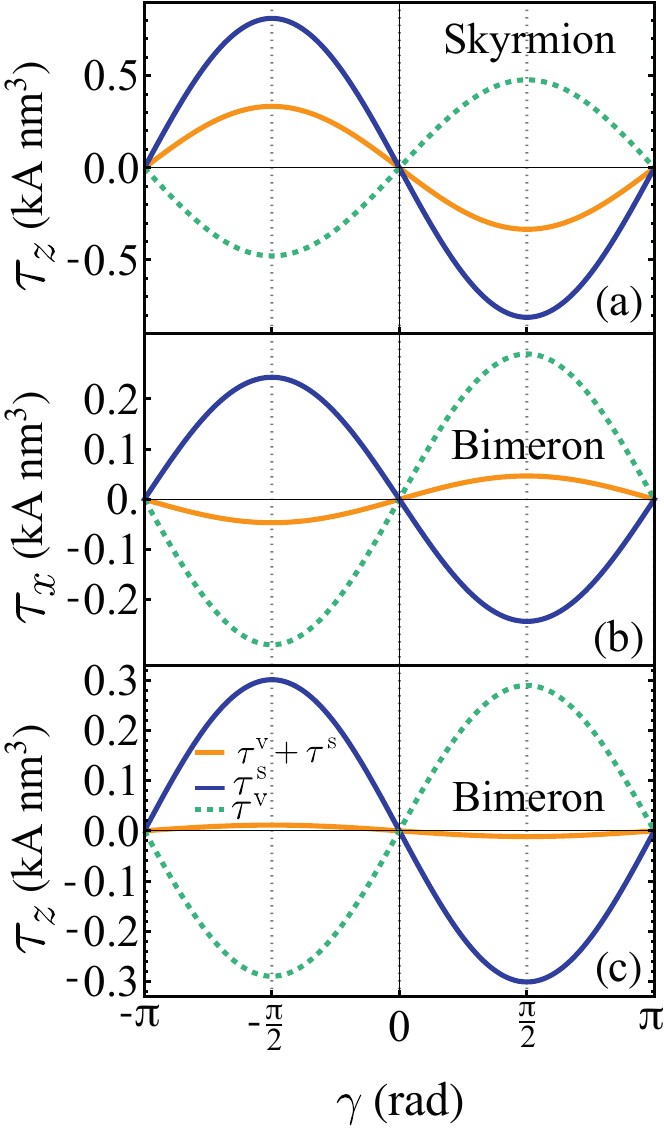}
	\caption{Relevant components of the volume toroidal moment as a function of the magnetic helicity $\gamma$ for different textures: (a) skyrmion ($z-$component) and (b,c) bimeron ($x-$ and $z-$components) for $R=100$ nm, $d=1$ nm, $M_{\rm{s}} = 658$ kA/m and $\lambda=5R/4$.}
	\label{fig:skyrmion_bimeron_gamma}
\end{figure}

The chosen model for the skyrmionic textures is given by $\mathbf{M}=M_{\rm{s}}\big[\sin\theta(\mathbf{r})\cos(\phi+\gamma)\,\hat{x}+\sin\theta(\mathbf{r})\sin(\phi+\gamma)\,\hat{y}+\cos\theta(\mathbf{r})\,\hat{z} \big]$ \cite{Bhowal2022}, where $\theta(\mathbf{r}) = \pi(1-r/\lambda)$, with $\lambda$ a characteristic radius for which $\mathbf{M}(r=\lambda)=-\mathbf{M}(0)$, $0\leq\phi\leq 2\pi$ and $\gamma$ is the helicity: $\gamma = 0\,(\pi/2)$ corresponds to a Néel$\,$(Bloch) skyrmion. The same expression can describe the antiskyrmion, by making $\phi \rightarrow -\phi$, while the bimeron texture is obtained by considering a $y-$axis rotation in $90^{\circ}$, which implies $M_{x}\rightarrow M_{z}$, $M_{y}\rightarrow M_{y}$ and $M_{z}\rightarrow -M_{x}$. From this, it is possible to model Skyrmioniums or $\ell \pi-$Skyrmions ($\ell=1,2,3...$) and $R = \ell\lambda$, which are skyrmion-like states stabilized in a ferromagnetic nanodisk of radius $R$ and thickness $d$. They are composed of a combination of skyrmions or multiple full spin rotations along the diameter with a different topological charge \cite{Pepper2018,Cortes2019,Mehmood2021,Ponsudana2021}. We also considered a linear meron model \cite{Radaelli2020}, given by $\mathbf{M} = M_{\rm{s}}\big[-\sin\theta(r)\sin\phi\,\hat{x}+ \sin\theta(r)\cos\phi\,\hat{y} + \cos\theta(r)\,\hat{z}\big]$, where $\theta = \frac{\pi r}{2R}$ for $r\leq R$. Table \ref{tab:table2} summarizes the volumetric and surface toroidal moments with their contributing faces and the Fig.~\ref{fig:skyrmion_bimeron_gamma} shows the dependence of the relevant components of $\boldsymbol{\tau}^{\rm{v}}$ with the helicity ($\gamma$). It can bee seen from the Table \ref{tab:table2} and the Fig.~\ref{fig:skyrmion_bimeron_gamma} that for $\gamma=0$, $\boldsymbol{\tau}^{\rm v}$ is always zero, then Néel-like textures do not induce dipolar nonreciprocity. Note that full skyrmioniums are composed of integer values of $\ell$; however, a non-integer value was chosen in Fig.~\ref{fig:skyrmion_bimeron_gamma} for instructive reasons, since for integer values $\boldsymbol{\tau}^{\rm s}$ is always zero for bimerons.

\begin{table}[htb]
\fontsize{10pt}{10pt}
\caption{Surface and volume toroidal moments for Skyrmion, bimeron, and meron. For antiskyrmion both toroidal moments are always zero. The brackets show the pairs of surfaces contributing to $\boldsymbol{\tau}^{\rm{s}}$.}
\renewcommand{\arraystretch}{1.25} 
\begin{ruledtabular}
\begin{tabular}{cccc}
& \textbf{Skyrmion} & \textbf{Bimeron} & \textbf{Meron}  \\
\colrule 
$\boldsymbol{\tau}^{\rm{s}}$ &  $\mathcal{C^{\rm{s}}}(0,0,\sin\gamma)|\{S_{z}\}$ &  $(\mathcal{D}^{\rm{s}}_1\sin\gamma,\mathcal{D}^{\rm{s}}_1\cos\gamma,\mathcal{D}^{\rm{s}}_2\sin\gamma)|\{S_{z}\}$  &  $(0,0,\mathcal{E^{\rm{s}}})|\{S_{z},S_{\rho}\}$   \\
$\boldsymbol{\tau}^{\rm{v}}$ & $\mathcal{C^{\rm{v}}}(0,0,\sin\gamma)$  & $\mathcal{D}^{\rm{v}}(\sin\gamma,\cos\gamma,\sin\gamma)$   & $(0,0,\mathcal{E^{\rm{v}}})$  \\
\end{tabular}
\end{ruledtabular}
\label{tab:table2}
\end{table}

The corresponding expressions that appears in Table \ref{tab:table2} are
\begin{align}
	\mathcal{C}^{\rm{s}} &= - M_{\rm{s}} d \frac{\pi  R \left[\pi ^2 d^2+24 \left(\lambda ^2+\pi ^2 R^2\right)\right] \sin \left(\frac{\pi  R}{\lambda }\right)-24 \lambda ^3+12 \lambda  \left(2 \lambda ^2-\pi ^2 R^2\right) \cos \left(\frac{\pi  R}{\lambda }\right)}{60 \pi ^2},\\
	\mathcal{C}^{\rm{v}} &= M_{\rm{s}} d \lambda  \frac{\left(2 \lambda ^2-\pi ^2 R^2\right) \cos \left(\frac{\pi  R}{\lambda }\right)-2 \lambda  \left[\lambda -\pi  R \sin \left(\frac{\pi  R}{\lambda }\right)\right]}{\pi ^2} ,\\
	\mathcal{D}^{\rm{s}}_1 &= -M_{\rm{s}}\pi R d (d^2 + 12 R^2)\sin\left(\frac{\pi R}{\lambda}\right) ,\\
	\mathcal{D}^{\rm{s}}_2 &= -M_{\rm{s}} d \frac{\pi  R \left[\pi ^2 d^2+24 \left(\lambda ^2+\pi ^2 R^2\right)\right] \sin \left(\frac{\pi  R}{\lambda }\right)-24 \lambda ^3+12 \lambda  \left(2 \lambda ^2-\pi ^2 R^2\right) \cos \left(\frac{\pi  R}{\lambda }\right)}{120 \pi ^2} ,\\
	\mathcal{D}^{\rm{v}} &= M_{\rm{s}} d \lambda  \frac{\left(2 \lambda ^2-\pi ^2 R^2\right) \cos \left(\frac{\pi  R}{\lambda }\right)-2 \lambda  \left[\lambda -\pi  R \sin \left(\frac{\pi  R}{\lambda }\right)\right]}{2 \pi ^2} ,\\
	\mathcal{E}^{\rm{s}} &= -M_{\rm{s}}d R\frac{\pi ^3 d^2+24R^2 \left(\pi^3 +4 \pi -8\right) }{60 \pi ^2} ,\\
	\mathcal{E}^{\rm{v}} &= M_{\rm{s}}d  R^3\frac{8 (\pi -2) }{\pi ^2}
\end{align}

\subsection{Toroidal moment of conical-helix texture on thin tubes}
By considering the following model for a conical-helix magnetization distributed over a thin tubular shell \cite{Mimica24},
\begin{align}
    	\frac{\mathbf{M}}{M_{\rm s}}  &= \cos(n\phi+q_{z} z + \psi)\sin\theta \hat{\rho} + \sin(n\phi+q_{z} z + \psi)\sin\theta \hat{\phi} + \cos\theta\hat{z} 
\end{align}
where $n$ is the azimuthal index which quantizes the azimuthal component of the pitch vector, $q_z$ is the $z-$component of the pitch vector, $\psi$ is a phase and $\theta$ is the cone angle.

the volume toroidal moment from Eq. \ref{eq:tau_v} gives
\begin{align}
\boldsymbol{\tau}^{\rm{v}} = \frac{1}{2} \int{dV\,      \{-z \sin\theta \sin (q_{z} z+n \phi +\psi )\hat{\rho} + [z \sin \theta  \cos (q_{z} z+n \phi +\psi )- \rho \cos \theta]\hat{\phi} + \rho \sin \theta \sin (q_{z} z+n \phi +\psi) \hat{z} \}     }.
\end{align}
By changing from cylindrical to Cartesian unitary vectors, i.e., $(\hat{\rho},\hat{\phi},\hat{z}) \rightarrow (\hat{x},\hat{y},\hat{z})$, and integrating in cylindrical coordinates for a thin tube with inner radius $\beta R$, outer radius $R$ and length $L$, $\boldsymbol{\tau}^{\rm v}$ becomes
\begin{align}
\boldsymbol{\tau}^{\rm{v}} =   M_{\rm s} R^2 \sin\theta \sin (n\pi) \bigg\{    &\frac{\left(1- \beta ^2\right)  \cos (n\pi+\psi ) }{2 q_{z}^2 (n+1)} \left[q_{z} L \cos \left(\frac{q_{z} L}{2}\right)-2 \sin \left(\frac{q_{z} L}{2}\right)\right] \hat{x} \nonumber \\
&+\frac{\left(1- \beta ^2\right) \sin (n\pi+\psi )}{2 q_{z}^2 (n+1)} \left[q_{z} L \cos \left(\frac{q_{z} L}{2}\right)-2 \sin \left(\frac{q_{z} L}{2}\right)\right] \hat{y} \nonumber  \\
&+\frac{2 \left(1- \beta ^3\right) R}{3 q_{z} n} \sin \left(\frac{q_{z} L}{2}\right) \sin (n\pi+\psi ) \hat{z} \bigg\}.
\label{eq:tv_conical_helix_tube}
\end{align}
By inspecting the $q_{z}$ and $n$ values it is found that this expression is only nonzero for $n=0$ and $n=-1$ and the result can be expressed as
\begin{align}
	\boldsymbol{\tau}^{\rm{v}} = &M_{\rm s}\frac{\,2 \pi  \left(1- \beta ^3\right)  R^3}{3 q_{z}} \sin \theta \sin \psi \sin \left(\frac{q_{z} L}{2}\right) \delta_{n,0} \hat{z} \nonumber \\
	&+ M_{\rm s} \frac{ \pi R^2 (1-\beta^2)}{2 q_{z}^2} \sin\theta \left[q_{z} L \cos \left(\frac{q_{z} L}{2}\right)-2 \sin \left(\frac{q_{z} L}{2}\right)\right]  \delta_{n,-1} ( \cos \psi \hat{x} + \sin\psi \hat{y}).
\end{align}
where the symbol $\delta_{i,j}$ denotes the Kronecker delta function. For the particular case $q_{z}=0$ and $n=0$,
\begin{align}
	\boldsymbol{\tau}^{\rm{v}} = M_{\rm s} \frac{  L R^3 }{3} \pi  \left(1 - \beta ^3\right) \sin \theta \sin \psi \, \hat{z}.
\end{align}
On the other hand, the toroidal moment $\boldsymbol{\tau}$ (Eq. \ref{eq:dubovik}) for the same texture is given by,
\begin{align}
    \boldsymbol{\tau} = &\frac{M_{\rm s} \pi  (\beta -1)  R \sin\theta \sin\psi}{30 q_z^3} \bigg[\sin \left(\frac{q_z L}{2}\right) \left(q_z^2 \left(3 L^2+4 \left(\beta ^2+\beta +1\right) R^2\right)-24\right)\nonumber\\
    &\hspace{8cm}+2 q_z L \cos \left(\frac{q_z L}{2}\right) \left(\left(\beta ^2+\beta +1\right) q_z^2 R^2+6\right)\bigg]\hat{z} \; \delta_{n,0} \nonumber \\
    &-\frac{M_{\rm s} \pi  (\beta^2-1)  R^2 \sin\theta}{20 q_z^2} \left[\sin \left(\frac{q_z L}{2}\right) \left(q_z^2 \left(2 L^2+3 \left(\beta ^2+1\right) R^2\right)-16\right)+8 q_z L \cos \left(\frac{q_z L}{2}\right)\right] (\cos\psi\,\hat{x} +\sin\,\psi \hat{y})\;\delta_{n,-1} \nonumber \\
    &+\frac{M_{\rm s} \pi  (\beta ^2-1) R^2 \sin\theta}{20 q_z^2} \left[\sin \left(\frac{q_z L}{2}\right) \left(\left(\beta ^2+1\right) q_z^2 R^2-4\right)+2 q_z L \cos \left(\frac{q_z L}{2}\right)\right](\cos\psi\,\hat{x} - \sin\psi\,\hat{y})\delta_{n,1}.
    \label{eq:tm_total_ch_cylindrical}
\end{align}
This last expression is only nonzero for $n=-1$, $n=0$ and $n=1$. Analogously, the surface toroidal moment $\boldsymbol{\tau}^{\rm s}$ for this texture using the Eq. \ref{eq:tau_s} is 

\begin{align}
    \boldsymbol{\tau}^{\rm s} = &\frac{M_{\rm s} \pi  (\beta -1) R \sin \theta \sin \psi}{30 q_z^3}\bigg[3 \sin \left(\frac{q_z L}{2}\right) \left(q_z^2 \left(L^2+8 \left(\beta ^2+\beta +1\right) R^2\right)-8\right)\nonumber \\
    &\hspace{7cm}+2 q_z L \cos \left(\frac{q_z L}{2}\right) \left(\left(\beta ^2+\beta +1\right) q_z^2 R^2+6\right)\bigg]\hat{z}\,\delta_{n,0} \nonumber \\
    &-\frac{M_{\rm s} \pi  \left(\beta ^2-1\right) R^2 \sin \theta  \cos \psi}{20 q_z^2} \left[\sin \left(\frac{q_z L}{2}\right) \left(q_z^2 \left(2 L^2+3 \left(\beta ^2+1\right) R^2\right)+4\right)-2 q_z L \cos \left(\frac{q_z L}{2}\right)\right]\nonumber\\
    &\hspace{11cm}\times(\cos\psi\,\hat{x} +\sin\,\psi \hat{y})\;\delta_{n,-1} \nonumber \\
    &+\frac{M_{\rm s} \pi  (\beta ^2-1) R^2 \sin\theta}{20 q_z^2} \left[\sin \left(\frac{q_z L}{2}\right) \left(\left(\beta ^2+1\right) q_z^2 R^2-4\right)+2 q_z L \cos \left(\frac{q_z L}{2}\right)\right](\cos\psi\,\hat{x} - \sin\psi\,\hat{y})\delta_{n,1}.
\end{align}
which, as \eqref{eq:tm_total_ch_cylindrical}, is only non-zero for $n=0$, $n=-1$ and $n=1$. It can be noticed that the terms associated with $n=1$ are exactly the same for $\boldsymbol{\tau}$ and $\boldsymbol{\tau}^{\rm s}$ and because of that, those terms do not appear for $\boldsymbol{\tau}^{\rm v}$.


\providecommand{\noopsort}[1]{}\providecommand{\singleletter}[1]{#1}%